\documentclass[olinenumbers, twocolumn]{aastex631}
\usepackage{xspace}
\usepackage{xcolor}
\definecolor{myblue}{RGB}{0, 0, 180}

\gdef\deg{$^{\circ}$}

\def\etal{\hbox{et al.}}

\gdef\eg{{\it e.g.,\ }}

\gdef\ltsima{$\scriptscriptstyle \; \buildrel < \over \sim \;$}
\gdef\simlt{\lower.3ex\hbox{\ltsima}}

\gdef\gtsima{$\scriptscriptstyle \; \buildrel > \over \sim \;$}
\gdef\simgt{\lower.3ex\hbox{\gtsima}}

\gdef\about{\raise.3ex\hbox{$\scriptscriptstyle \sim $}}

\def\gs{\mathrel{\raise0.35ex\hbox{$\scriptstyle >$}\kern-0.6em 
\lower0.40ex\hbox{{$\scriptstyle \sim$}}}}
\def\ls{\mathrel{\raise0.35ex\hbox{$\scriptstyle <$}\kern-0.6em 
\lower0.40ex\hbox{{$\scriptstyle \sim$}}}}

\def\etal{\hbox{et al.}}

\def\OIII{\hbox{[O III]}$\,\,$}
\def\Ha{\hbox{H$\alpha$}$\,$}

\def\Msun{\rm{\hbox{$\,$M$_{\odot}$} }}				
\def\Msuns{\rm{\hbox{$\,$M$_{\odot}$}}}	
\def\MsunYr{\rm{\hbox{M$_{\odot}$} yr$^{-1}$}}           

\def\24m{\hbox{24\,$\micron$}$\,$}
\def\10-18{\hbox{$\times~10^{-18}$}}
\def\deg{\hbox{$^{\circ}$}}
\def\SEDz{\hbox{\emph{SEDz*} }}
\def\chisq{\hbox{${\chi}^2$} }
\def\JWST{\hbox{\emph{JWST} } }
\def\EZ{\hbox{$z\_a$}}
\def\zSED{\hbox{$z_{SED}$} }

\def\T0{{$t_0$}}

\defcitealias{Merlin2022}{Paper II}
\defcitealias{Leethochawalit2022}{Paper X}
\defcitealias{Nanayakkara2022}{Paper XVI}

\shorttitle{JADES-SEDs-Star formation histories}
\shortauthors{Dressler et al. }
\graphicspath{{./}{figures/}}

\begin{document}

\title{Building the First Galaxies --- Chapter 2. Starbursts Dominate The Star Formation Histories of 6$<$\MakeLowercase{z}$<$12 Galaxies}

\author[0000-0002-6317-0037]{Alan Dressler}
\affiliation{The Observatories, The Carnegie Institution for Science, 813 Santa Barbara St., Pasadena, CA 91101, USA}

\correspondingauthor{Alan Dressler}
\email{dressler@carnegiescience.edu}

\author[0000-0002-7893-6170]{Marcia Rieke}
\affiliation{Steward Observatory, University of Arizona, 933 N Cherry Ave, Tucson, AZ 85721, USA}

\author[0000-0002-2929-3121]{Daniel Eisenstein}
\affiliation{Center for Astrophysics  Harvard \& Smithsonian, 60 Garden Street, Cambridge, MA 02138, USA}

\author[0000-0001-6106-5172]{Daniel P. Stark}
\affiliation{Steward Observatory, University of Arizona, 933 N Cherry Ave, Tucson, AZ 85721, USA}

\author[0000-0002-6317-0037]{Chris Burns}
\affiliation{The Observatories, The Carnegie Institution for Science, 813 Santa Barbara St., Pasadena, CA 91101, USA}

\author[0000-0003-0883-2226]{Rachana Bhatawdekar}
\affiliation{European Space Agency (ESA), European Space Astronomy Centre (ESAC), Camino Bajo del Castillo s/n, 28692 Villanueva de la Cañada, Madrid, Spain; European Space Agency, ESA/ESTEC, Keplerlaan 1, 2201 AZ Noordwijk, NL}

\author[0000-0001-8470-7094]{Nina Bonaventura}
\affiliation{1. Cosmic Dawn Center (DAWN), Copenhagen, Denmark 2. Niels Bohr Institute, University of Copenhagen, Jagtvej 128, DK-2200, Copenhagen, Denmark 3.Steward Observatory University of Arizona 933 N. Cherry Avenue Tucson AZ 85721, USA}

\author[0000-0003-4109-304X]{Kristan Boyett}
\affiliation{School of Physics, University of Melbourne, Parkville 3010, VIC, Australia}
\affiliation{ARC Centre of Excellence for All Sky Astrophysics in 3 Dimensions (ASTRO 3D), Australia}

\author[0000-0002-8651-9879]{Andrew J. Bunker}
\affiliation{Department of Physics, University of Oxford, Denys Wilkinson Building, Keble Road, Oxford OX1 3RH, UK}

\author[0000-0002-6719-380X]{Stefano Carniani}
\affiliation{Scuola Normale Superiore, Piazza dei Cavalieri 7, I-56126 Pisa, Italy}

\author[0000-0003-3458-2275]{Stephane Charlot}
\affiliation{Sorbonne Universit\'e, CNRS, UMR 7095, Institut d'Astrophysique de Paris, 98 bis bd Arago, 75014 Paris, France}

\author[0000-0002-8543-761X]{Ryan Hausen}
\affiliation{Department of Physics and Astronomy, The Johns Hopkins University, 3400 N. Charles St., Baltimore, MD 21218}

\author{Karl Misselt}
\affiliation{Steward Observatory, University of Arizona, 933 N Cherry Ave, Tucson, AZ 85721, USA}

\author[0000-0002-8224-4505 ]{Sandro Tacchella}
\affiliation{Cavendish Laboratory, University of Cambridge, 19 JJ Thomson Avenue, Cambridge, CB3 0HE, UK}
\affiliation{Kavli Institute for Cosmology, University of Cambridge, Madingley Road, Cambridge, CB3 0HA, UK}

\author[0000-0001-9262-9997]{Christopher Willmer}
\affiliation{Steward Observatory, University of Arizona, 933 N Cherry Ave, Tucson, AZ 85721, USA}



\begin{abstract}

We use \SEDz --- a code designed to chart star formation histories (SFHs) of $6<z<12$ galaxies --- to analyze the SEDs of 894 galaxies with deep \emph{JWST/NIRCam} imaging by \emph{JADES} in the \emph{GOODS-S} field. We show how \SEDz matches observed SEDs using stellar-population templates, graphing the contribution of each epoch-by-epoch to confirm the robustness of the technique.  Very good SED fits for most SFHs demonstrates the compatibility of the templates with stars in the first galaxies --- as expected, because their light is primarily from main-sequence A-stars, free of post-main-sequence complexity and insensitive to heavy-element compositions. We confirm earlier results from \citet{Dressler2023}: (1) Four types of star formation histories: SFH1 --- burst; SFH2 --- stochastic; SFH3 --- `contiguous' (3-epochs); and SFH4 --- `continuous' (4-6 epochs); (2) Starbursts --- both single and multiple --- are predominate ($\sim$70\%) in this critical period of cosmic history, although longer SFHs (0.5-1.0 Gyr) contribute one-third of the accumulated stellar mass.   These 894 SFHs contribute $10^{11.14}$, $10^{11.09}$, $10^{11.00}$, and $10^{10.60}$ \Msun for SFH1-4, respectively, adding up to $\sim$4$\times10^{11}$\Msun by $z=6$ for this field.  We suggest that the absence of rising SFHs could be explained as an intense dust-enshrouded phase of star formation lasting tens of Myr that preceded each of the SFHs we measure.  We find no strong dependencies of SFH type with the large-scale environment, however, the discovery of a compact group of 30 galaxies, 11 of which had first star formation at $z=11-12$, suggests that long SFHs could dominate in rare, dense environments. 

\end{abstract}



\vspace{20mm}
\section{Star Formation Histories and the Birth of Galaxies --- \emph{JWST}'s ``Prime Mission''}
\label{sec:Prime mission}

The \emph{HST \& Beyond Committee} \citep{Dressler1986} chose a prime mission for the space telescope that would become the extrordinary \emph{JWST}.  The unanticipated reach of the `reborn' \emph{Hubble Space Telescope} to galaxies with redshifts $z=2-3$ --- only 2 billion years after ``the beginning" --- held extraordinary promise for learning how the modern universe actually \emph{began}, when first generations of stars collected into the first galaxies.  Identifying this as the key moment in our origins --- rather than, for example, the Big Bang --- hinges on the idea that the miracle of \emph{this universe}, and more to the point, of \emph{life}, is its incredible complexity.  The chemical variety and abundance that made this possible could never have built to a critical level without multiple generations of heavy-element-producing stars, residing in these giant reservoirs we know as \emph{galaxies}. 

\emph{``The Hubble"} as we now affectionately call it, took us to the trailhead of one of humanity's greatest journeys: traveling to this remote past to follow the story of our origins back to the present day.

Central to that mission is the simple notion that galaxies began in very different conditions  --- ``cosmic environments" --- at a time when dark matter was sculpting a terrain of mountains and valleys in a formerly smooth-as-glass universe.  There and then, ``ordinary matter" ---  the baryons that we would be made of --- were gathered and concentrated by gravity in a way that would fundamentally alter the very nature of the universe. We had asked questions: how fast, how turbulent, how explosive, how dynamic were the forces that shaped these galaxies, so different from our times, when a slow and irreversible decline dooms this universe to a grinding conclusion that we --- fortunately --- will never see.

The tremendous effort and challenge of building \emph{JWST} is behind us. The unearthly, spectacular views we now have of the universe --- near and far --- let us start that remarkable journey with a large, representative sample of faint, newborn galaxies, for a first look at how they survived their tumultuous births. 

Where to start? A long-standing and popular approach to the study of galaxy formation and development has been to push to higher and higher redshift to compare the properties of ever-younger galaxies. \emph{JWST} has already exceeded expectations with images and confirming spectroscopy of starforming galaxies at $z\sim13$ \citep{2023NatAs...7..611R, CurtisLake_2023}, a scant 300 Myr after the Big Bang.  Other studies identify candidates for even earlier galaxies; deeper searches and microlensing by rich clusters \citep{Treu2022a} promise to provide candidates as early as redshift z$\sim$18.  However, by their very nature, such searches find extraordinary objects --- the brightest sources at the earliest epochs.  In order to assemble a representative sample of the first galaxies, a later starting point is required: early \emph{JWST} results suggested z$\sim$12 as an epoch where dozens of galaxies could be identified in the \emph{JADES} \emph{GOODS-S} field, with numbers increasing rapidly into the many hundreds for z$\sim$6. This motivated the choice of $6<z<12$ for the present study of the star formation histories. But for their number, these are the ancestors of the common galaxies we find today.

The paper is organized as follows: Section~\ref{sec:special time} briefly reviews the  methodology --- the program \SEDz --- details of of which can be also found in \citet{Dressler2023}, hereafter, Paper 1. Section~\ref{sec:Deriving SFHs} lays the foundation for this study, showing and explaining a new format for \SEDz results, with new data that highlight the four types of SFHs.  Section~\ref{sec:Choosing a sample} explains how a robust, representative sample of $\sim$900 galaxies was chosen from a catalog of \emph{GOODS-S} sources with 9-band \emph{NIRCam} fluxes (SEDs) from the \emph{JWST Advanced Deep Extragalactic Survey} --- \emph{JADES}.~~  Section~\ref{sec:Examples SFHs} shows 72 examples of SFHs for four redshift ranges $6<z<12$, as measured by \emph{SEDz*}. {Section~\ref{sec:SFHs_Across_Time} looks for changes in the mix of SFHs over time, and  Section~\ref{sec:Galaxy born} presents a summary of the birth of stellar mass --- when, where, and how much --- from z$\sim$12 down to z$\sim$6.  Section~\ref{sec:SFHS Space} investigates correlations of SFHs with environment, and Section~\ref{sec:The Age of Starbursts} looks ahead to the implications of this study for future observational and theoretical programs.  Appendix A shows histograms of SFH type with local density and nearest-neighbor distance, Appendix B shows examples of SFH templates used by \SEDz, and Appendix C discusses the impact of emission lines on the continuum-dominated SEDs of this study.}

A standard cosmology with $\Omega_{\rm m}=0.3$ $\Omega_{\Lambda}=0.7$ and H$_0$=70 km s$^{-1}$ Mpc$^{-1}$ is adopted throughout. 

\section{A Special Tool for a Special Time}
\label{sec:special time}

The buildup of stellar mass in a galaxy is expressed as its {\it star formation history} (SFH), that is, the rate of star formation changing with time.  Measuring the star formation histories of galaxies has proven more than challenging, because while groups of young, massive stars are easy to recognize, populations of stars older than a billion years --- though still young compared to the universe --- age in such a way that it's not possible to tell a populatioh of old stars from another that is even older.

As part of a galaxy survey of 6 billion years of cosmic history, \citet{Kelson2014}\ developed a tool for analyzing a galaxy's ``spectral energy distribution"--- SED, basically, its rainbow of color. By measuring SEDs for galaxies observed $\sim$4 Gyr earlier in cosmic history, Kelson's analysis leveraged those observations to look back $\sim$2 Gyr further, to a time when the universe was only around half its present age.\footnote{...an epoch we can see by ``looking back in cosmic time" --- thanks to the late arrival of light in transit for billions of years.} Because of that, the part of a galaxy's history that \emph{could} be reliably measured was extended to when the universe had half its present age, in this way revealing a substantial fraction of a galaxy's growth history.  Applying this approach to a sample of a galaxies observed at redshift $z\sim0.4-0.8$, \citet{Dressler2018} discovered that $\sim$20\% of galaxies were still growing rapidly --- with \emph{increasing} star formation rates (SFRs), in an era that was thought to be a time of decline for \emph{all} galaxies as massive as the Milky Way.  

The motivation for this study, then, emerged directly from that one.  The ability to see only one billion years (or at most two) of star formation history, a small part of a mature galaxy's lifetime, stretches to ``longer than the age of the universe" during the period when \emph{JWST's} prime mission kicks in.  Our job has been to gather sufficiently good SEDs, using \emph{JWST's} near-infrared camera, \emph{NIRCam}, to make accurate measurements of galaxy brightness at a series of colors --- infrared light from 1-5\micron. Such data can fully constrain the ages of populations of stars that make up a galaxy, and lining up their stellar masses --- epoch by epoch --- constitutes a star formation history of how the galaxy was built.

In this paper we show the results of our second attempt (see Paper 1) to transform the flux measurements of each galaxy --- its SED --- into star formation histories, the buildup of stellar mass over the first billion years. Our subjects are a greatly-expanded sample of 894 galaxies at redshifts $6<z<12$ with 9-band near-infrared fluxes from \emph{JWST/NIRCam}.  We provide a short description of a program code --- \SEDz --- written expressly for the purpose of reconstructing the histories of galaxies from the rich information encoded in their SEDs. By choosing a sample of galaxies with ages less than billion years, their fast-evolving stellar populations will be recorded, for us to play back.

The development of \SEDz followed Kelson's  maximum-likelihood Python code for analyzing galaxy SEDs from the \emph{Carnegie-Spitzer-IMACS Survey} \citep{Kelson2014}, through the combination of stellar population templates.   The program effectively isolated light from younger ($\ls1$  Gyr) populations of {{main-sequence} $A$-stars and led to the discovery of ``late bloomers" --- the $\gs$20\% of galaxies at $z\sim0.5$ that produced at least 50\% of their stellar mass in the previous 1 Gyr, that is, \emph{rising} SFRs instead of the falling SFRs that are conventionally described as predominant after $z=1$.  These \emph{late bloomers} challenge theoretical models that tightly couple the growth of galaxies to that of their dark matter halos, because while it is possible to think of mechanisms that could delay star formation,  it is not easy to imagine why some halos might form so much later than all the rest.

With this approach, the program \SEDz was developed to exclusively measure SFHs of the first galaxies.  In Paper I we described the challenges associated with measuring SFHs for stellar populations with ages of more than 1 Gyr, a deficiency turned on its head when the the population under study has an age of $\tau\ls$1 Gyr.  \emph{SEDz*} takes advantage of this unique astrophysical opportunity that comes from the billion-year lifetimes of $A$-stars \citep{Dressler1983, Couch1987}.  Because they evolve rapidly over a Gyr, it is possible to \emph{derive} SFHs from SEDs for $A$-star-dominated populations, and vice-versa. $A$-stars are among the simplest main-sequence stars \citep{Morgan1973}: basically \emph{black-body} radiators with a relatively simple internal structure and an opacity produced by hydrogen absorption, thus free from the complications of chemical abundance and post-main sequence evolution.  These are critical and unique advantages for measuring SFHs during the first billion years.

The data input to \SEDz are SEDs --- flux measurements in \emph{NIRCam's} 7 wide bands.\footnote{F090W, F115W, F150W, F200W, F277W, F356W, \& F444W. Fluxes in the narrower bands F335M and F410M are not used in the SED fit, but are valuable in showing the presence of strong emission, particularly \OIII and \Ha\ at these redshifts.} \emph{SEDz*} uses a non-negative least squares (NNLS) engine \citep{Lawson1995} and custom star-formation templates \citep{Robertson2010}\footnote{The templates are based on \citet{Bruzual2003} models, but include emission  and non-stellar continuum from star forming regions in calculating the fluxes.} that are essentially a set of vectors which have a significant amount of ``orthnomality," as can be seen in the plots in the \emph{Appendix B}, where a more complete discussion of templates can be found.

For this study, \SEDz divides redshifts $6<z<12$ into 7 integer steps, $12, 11...6$ lasting 42, 50, 61, 79, 101, 134, 186 Myr, respectively. \SEDz operates with two sets of SED templates, one with 10 Myr bursts (unresolved after subsequent star formation) and another characterized by constant star formation (CSF). The program builds up a SED by adding stellar population templates (starting at $z=12$ and working down) as needed to improve the fit, and evolving them forward --- adding up subsequent populations to improve the fit, as measured by ${\chi}^2$. The epoch of observation (OE) is chosen as the lowest \chisq and the star formation history is read off as the stellar mass contributed by each scaled template that, in combination, make the best fit.  This can include the addition of a CSF template which signals constant star formation at OE.  \SEDz can combine a CSF template with a final burst template to expand its fitting possibilities, as shown in \emph{Appendix B}.  

{\emph{SEDz*} requires no ``priors" in the important sense that every trial solution, as lower-redshift templates are added one-by-one, is fully independent of previous ones.  This allows for modeling free of any particular functional form (\eg{non-parametric}) and allowing for a range of galaxy histories.  However, that range is bounded: only 33 templates are used (28 burst, z=12 to z=6, evolving, and 5 CSF, z=10, 9, 8, 7, 6 --- at the epoch of observation). Since flux ratios are fixed for each template, choosing a template constrains all 7 bands --- there is no fitting of individual bands, so an acceptable solution is not guaranteed: finding one indicates also that the SED \emph{can} be well-described by the templates.}\footnote{Since the templates come from present- epoch stellar populations, suitability to very-high-redshift stellar populations was unknown.}

Paper I describes the working of the code in considerable detail, so we do not repeat it here.  Also discussed in Paper I are tests of \SEDz, including its ability to reproduce the SEDs of synthesized galaxies in a simulated deep field, from the \emph{NIRCam} ``Data Challenge" (Williams+2018). The \emph{Appendix} of Paper 1 shows how test-SFHs generated by combining the stellar population templates were recovered by \SEDz, and how the distinction between bursts and extended SFHs is robust.

As in Paper 1, we neglect the potential impact of dust.  We note, as before, that galaxies in our sample seem to be well described by SEDs with little or no dust, consistent with the results of several papers showing that these initial \emph{JWST}-selected samples are uniformly blue \citep{Nanayakkara2022} and fairly dust free. {Our library of stellar population templates includes sets with extinctions of $A_v$ = 0.4 \& 1.0.  We show in Figure~\ref{CSFandCSF+bursts_templates} that extinction at the level of $A_V$ = 0.4 is ruled out, in fact, even $A_V$ = 0.1 would produce a reddened SED not found in the 894-galaxy sample.  While we did find 27 other examples of steeply rising SEDs in the current work, their identification as galaxies at high redshift is not possible with \SEDz, because of its reliance on the A-star model.\footnote{If \SEDz was applied with the highly reddened A-stars of the dusty templates shown in Figure~\ref{CSFandCSF+bursts_templates}, the variation caused by unknown amounts of dust would overwhelm the variation that could be attributed to different ages of stars.} We conclude from this that our sample is by construction a ``no-extinction" sample, and that it does describe more than 95\% of galaxies found by deep \emph{NIRCam} imaging in this field.}

However, we do speculate here that the absence of rising SFHs in this and our previous study could indicate that this phase of galaxy building is largely hidden by dust, and that the SFHs we find began with 10--20 Myr of intense, dust-enshrouded star formation, 
then cleared by its explosive feedback.

\section{Deriving SFHs of the First Galaxies with SEDz}
\label{sec:Deriving SFHs} 

In Paper 1, we introduced and explained the \SEDz code and applied it to data from the \emph{GLASS} \emph{JWST/ERO} program  \citep{2022ApJ...935..110T} from  parallel imaging with \emph{NIRCam} of  \emph{NIRISS} grism spectroscopy of the cluster Abell 2744 \citep{Merlin2022}. Some challenges in processing and calibrating one of the first deep-imaging programs --- a crucial rationale for the \emph{ERO} program --- limited the targets of Paper 1 to only 24 galaxies that were judged suitable for a first measurement of SFHs. 

To that point \SEDz had only been tested on simulated data of the \emph{NIRCam} deep imaging program (the \emph{Data Challenge} \citep{Williams2018}, and on simulated SFHs produced using \SEDz itself, an admittedly easier test to pass. Valid questions had been raised about how different the SEDs of the earliest galaxies might be from their descendants, particularly because the nature of stellar populations at these early times was largely unknown.  However, the application of \SEDz to the first such data produced encouraging, and surprising results, in the sense that the program was able to reproduce 24 complex 9-band SEDs with the code's limited library of stellar population templates.\footnote{In fact, our first attempt to apply \SEDz to an early photometric catalog, which expressed fluxes in micro-Jy instead of the expected nano-Jy, produced only the SED-equivalent of gibberish.}  In other words, SFHs with recognizable characteristics, bursts --- single and multiple, and multi-epoch star formation, with reasonable masses in the range of $10^8$ to $10^9$ \Msuns, fit all 24 SEDs within the errors of the photometry. Considering the limited number of stellar population templates available to \SEDz, and their unique shapes, obtaining good fits from the `get-go' was surprising, and remarkable.

With the comparatively exquisite photometry from \emph{NIRCam} imaging of \emph{GOODS-S} for \emph{JADES} in late 2022, deeper imaging data have led to a $\sim$900-galaxy sample at redshifts $6<z<12$, and our analysis of these new data confirms the basic conclusions of Paper 1: good reproduction of observed SEDs and four SFH `types' and, in particular, well-establishing the unexpected prominence of starburst SFHs over longer, steadier runs of star formation.  Moreover, we now have sufficient data to begin to examine the dependence of these SFHs with redshift, mass, environment, and large-scale structure.

We begin by revisiting the variation in SFH type. Figure~\ref{fig:4SFHs} introduces a new data format for \SEDz output and gives examples of the four types of SFHs found in Paper 1 and now in this paper, demonstrating both how the code derives SFHs and that it does so with considerable fidelity.  The left-hand box displays the observed SED --- fluxes in each of the 9 bands with 1$\sigma$ error bars.  The NNLS solution \SEDz found by combining stellar population templates is the magenta band --- the quartile range of 21 iterations, each a random perturbation of the SED by its errors. The \chisq of the fit is inset in the upper-left: a prominent minimum in \chisq defines the observed redshift, or OE. The solution and operation of \SEDz is recorded in the right box, which shows the stellar mass added at each of 7 epochs (integer redshifts z = 12, 11...6).  This is the how the SFH is calculated, by scaling and combining stellar population templates to make the best NNLS fit to the observed SED.

{In Figure~\ref{fig:4SFHs}, this `best fit' for the upper-left SED is a starburst --- a single epoch of star formation observed at $z=7$, in this case, a combination of a burst and constant star formation --- CSF.  As shown in the left hand panel of Figure~\ref{CSFandCSF+bursts_templates} (\emph{Appendix B}), the slope of CSF alone is more level, and a burst alone much steeper, than what is observed.  \SEDz has determined that at a ratio of 2:1 (more CSF than burst) is an excellent fit to the observed SED.  That is, this two-component stellar population model from present-epoch stars in our Galaxy ``works." While it might seem remarkable that present epoch stellar populations provide an almost perfect fit, in one sense this is completely unremarkable: the $A$-stars that dominate the light of this starburst at the OE are the least-complicated stars along the main sequence: a fully convective core and a fully radiative envelope with opacity from hydrogen ions --- no metals required!}

\begin{figure*}[t!]
\includegraphics[width=.5\textwidth]{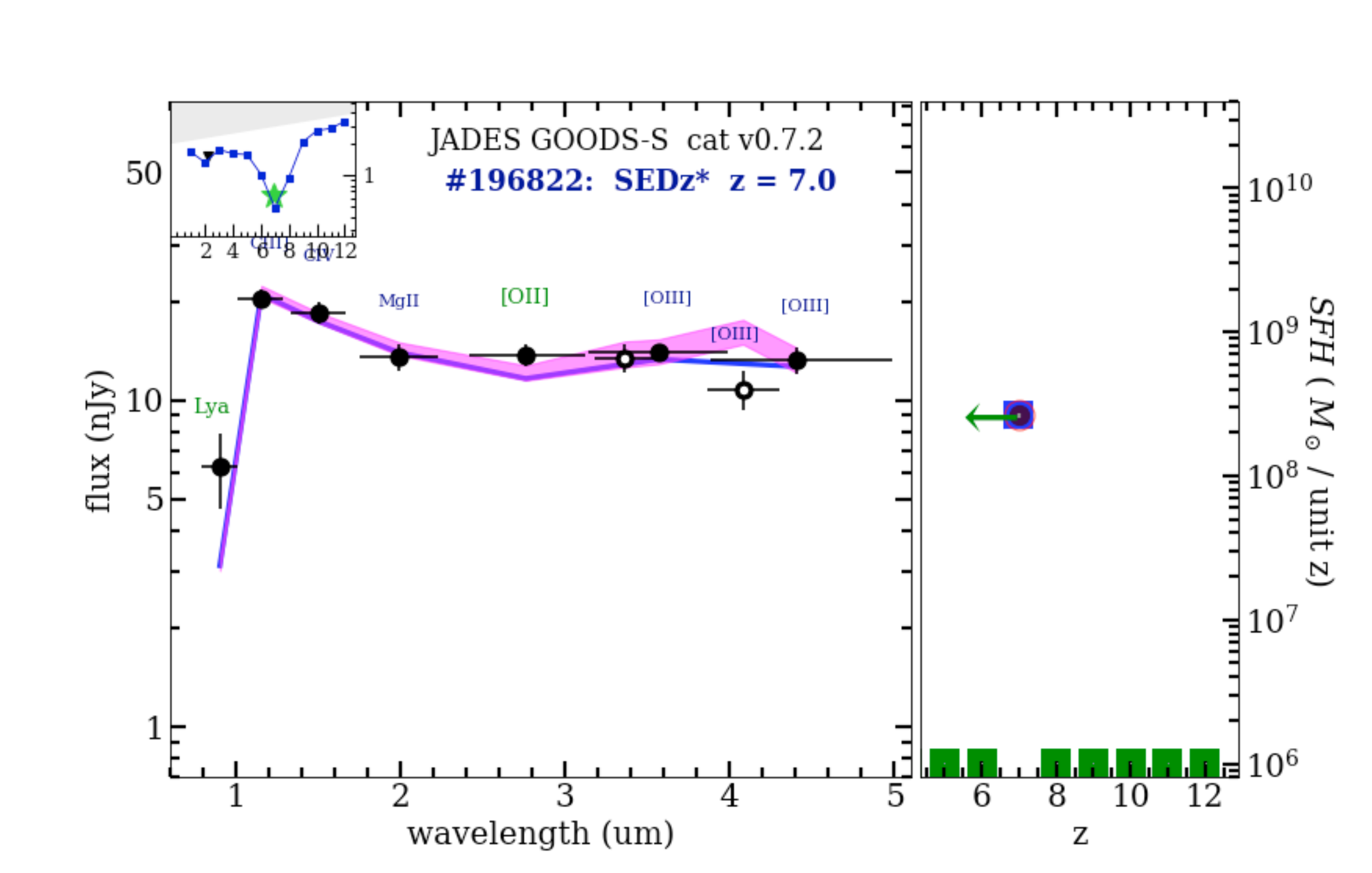}%
\includegraphics[width=.5\textwidth]{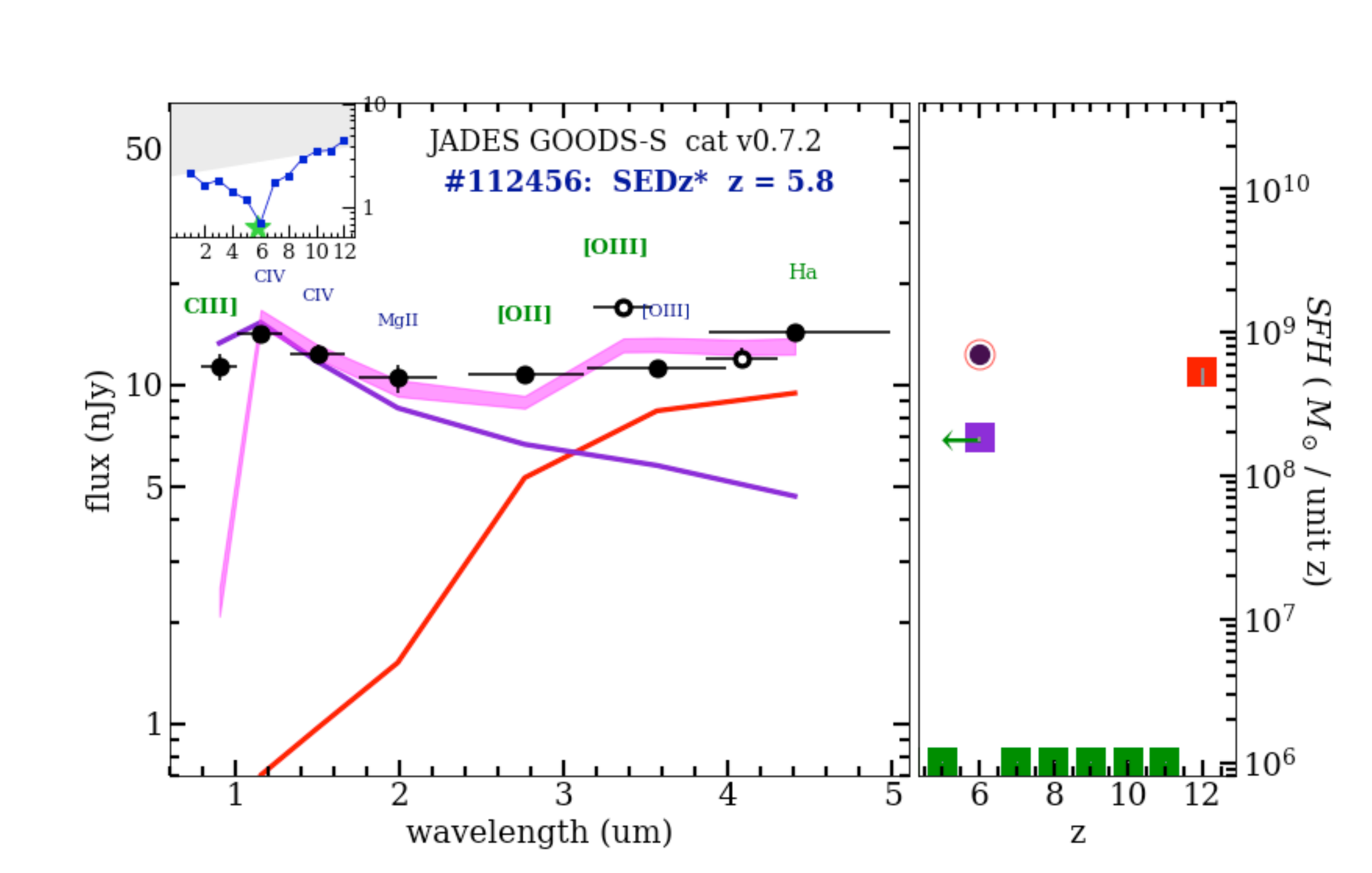}
\includegraphics[width=.5\textwidth]{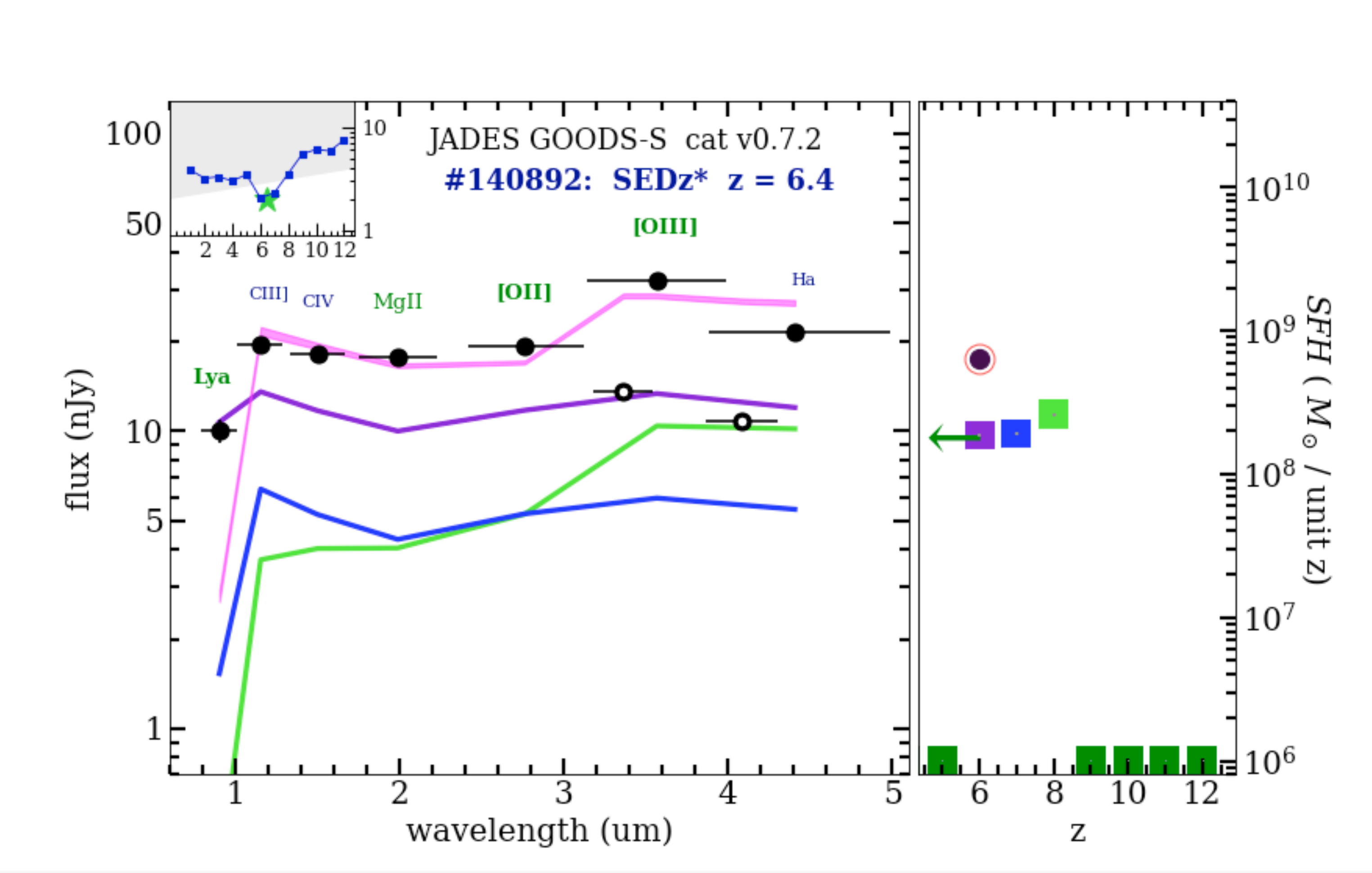}%
\includegraphics[width=.5\textwidth]{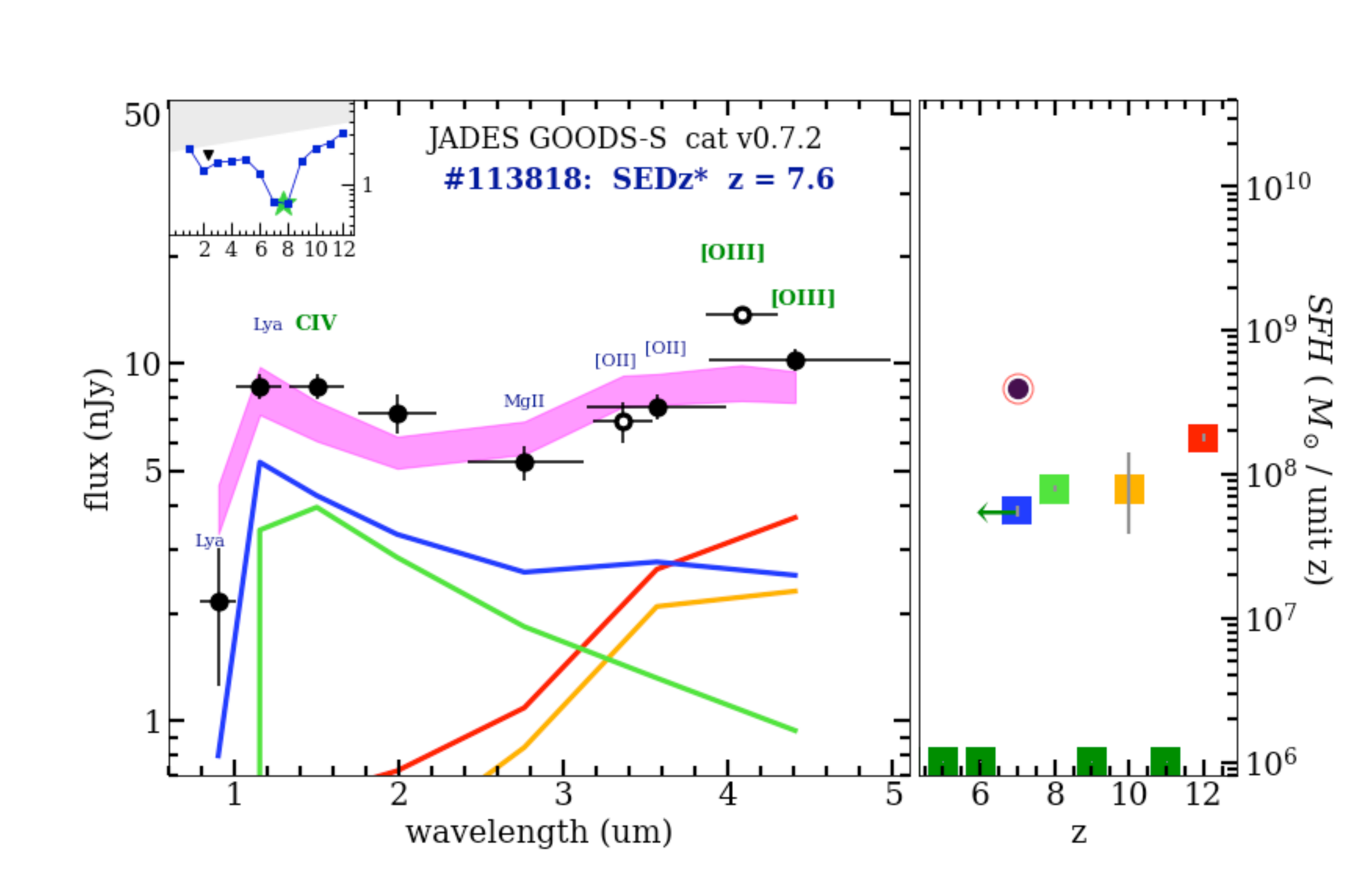}
\caption{Examples of the four types of SFHs $5<z<12$ found with \SEDz. The text describes how, considered together, these examples demonstrate the fidelity of \emph{SEDz*}-derived SFHs. The left box shows the observed SED (black points with error bars) from 9-band \emph{NIRCam} imaging and the NNLS fit of \SEDz --- the magenta band, showing the `quartile range' of 21 trials, with all data points perturbed by 1$\sigma$ random errors. The run of \chisq is shown in the insert at upper left; the dip (green star) marks the observed redshift. Most important is the SFH corresponding to this best fit, shown in the right box. Each epoch records the stellar mass from that epoch's star formation as a small colored box. For each epoch that contributes to the SED fit, the flux contributed to the solution is plotted on the left box, as a line of the same color (below the observed SED). Error bars, based on the quartile ranges of the \SEDz fit, are typically smaller than the boxes, but a prominent exception is the $z=10$ burst in the lower-right SFH whose contribution to the mass is uncertain within a factor of $\sim$5 and may or may not contribute significantly.  Error bars do not include systematic errors, such as errors associated with photometry at these faint levels, but these are unlikely to perturb the \emph{shape} of the \SEDz solution. While error bars corresponding to factors-of-two uncertainty in mass are common in the large sample of the present work, such errors are not large enough to admit distinctly different SFHs. The four panels show a starburst (SFH1 --- upper-left)  and a double burst (SFH2 --- upper-right), a `short' but contiguous run of star formation   (SFH3 --- lower-left) and a longer, continuous SFH (SFH4 --- lower right) covering half of the first $\sim$800 Myr of cosmic history to that point. The expected locations of prominent emission lines are shown above the SED, in blue; a larger green font marks a possible detection as excess flux compared the best fit to the flux of continuum \emph{with} emission lines.
}
\label{fig:4SFHs}
\end{figure*}

{The stellar mass calculated for star formation at $z=7$ is $\sim$3$\times10^8\Msuns$. The green arrow signifies the CSF  contribution, with ongoing star formation at OE.\footnote{A small arrow indicates that less than half of the light comes from the CSF template (no arrow == none), and a larger arrow means more than half, and as much as `all.'}  The CSF template accounts for emission at OE, detected as the modest elevation of F356W and F444W from the stellar continuum level (F410M) due to \OIII\ emission.  The CSF template only serves at OE, because stars formed in a prior epoch are, by definition, old.  In fact, although a burst or CSF is the way \SEDz models the growth of in stellar mass through the epochs, these are indistinguishable by the next epoch, becoming an ``old" population, regardless of how the mass was spread over the interval. It is also important to remember that, despite representing continuing CSF at OE, most of the light from the CSF template is from young A-stars --- the O \& B stars, and any non-stellar continuum and emission are gone, except for the last 10-20 Myr.}

This starburst example begs the question, could another SFH --- for example, a combination of bursts --- also reproduce the top-left SED?  The top-right panel shows that the answer is `no,' by means of a SED that is perfectly fit with a combination of \emph{two} bursts: one of $\sim$5$\times10^8$\Msun at $z\approx12$ and a $\sim$2$\times$10$^8$\Msun CSF+burst $\sim$500 Myr later (OE z$\approx$6). The integrated mass of this SFH is the red-encircled black dot at $\sim$7$\times$10$^8$\Msuns. Epochs without star formation are marked as lower limits of $10^6$\Msuns. In this study, at the depth of the $JADES$ data for the \emph{GOODS-S} field, detections as low as $10^7$\Msun have been made, but are incomplete below $\sim$5$\times10^7$\Msun (limited by the `noise' from other epochs of star formation), and severely so below $2\times10^7$\Msun.  Nevertheless, it is clear that multiple epochs of star formation cannot combine to produce the SED on the left, nor can a single epoch of star formation imitate a double.

{Perhaps the most revealing feature of this double-burst example is how clearly it shows \SEDz combining two widely-separated epochs of star formation to match the observed SED.  The two colored curves below the SED are A-star templates that correspond to the age of the two bursts.  The $z=12$ burst in orange is that of an `old' stellar population ($\sim$500 Myr before OE), while the purple curve has the strong ultraviolet flux of a young population ($\sim$100 Myr), demonstrating the ability of \SEDz to resolve $6<z<12$ epoch into old and young populations. This time it is the burst at $z=6$ that dominates CSF, by 2:1, which accounts for the steeper slope of the purple template, as shown in the right plot in Figure~\ref{CSFandCSF+bursts_templates} (Appendix).  It is the striking difference between the two templates that together reproduce the observed SED that shows how accurately \SEDz can decompose a SED into separate stellar populations, and how well a SFH can be matched with a comparatively small number of A-star templates.}\footnote{As in the single-burst case, this double burst shows signs of star formation (elevation of the medium bands) that is contributed by CSF at $z\sim$6.  As we discuss briefly in Section~\ref{sec:Examples SFHs}, and show in detail in \emph{Appendix C}, this emission is not responsible for the rise in red flux that is provided by star formation at $z=12$.  That contribution has already been accounted for in predicting the two broad-band fluxes, F356W and F444W.}

{Before moving on to examples of longer SFHs, we want to clarify why these two examples are called `starbursts,' given that each represent a few hundred million solar masses in approximately one hundred Myr.  An averaged SFR of only a few \MsunYr} {is not usually considered a `burst.'  However, we have no information on the \emph{history} of star formation (SFRs) over these individual epochs: we're not measuring SFRs, but total stellar mass associated with that epoch.  What makes the top left example a `burst' is that the $z=7$ epoch contains only young A-stars --- that is the template that fits the SED. \emph{No older A-stars --- stars born in previous epochs --- are detected}.  As we discuss in Section~\ref{sec:The Age of Starbursts}, we think it  more likely that the stellar mass we find within a single epoch was born in $\sim$10-20 Myr within the epoch, with SFRs in the `tens'} \MsunYr), {and a falling SFR (or even a shutdown) in the $50-100$ Myr after. Such `more-likely' SFHs would account for the a A-star dominated SED and a corresponding absence of older A-stars, $\tau\gs$150 Myr. And that is a `burst', by any definition.}

The triple-epoch SFH shown in the bottom-left panel is common in the sample, particularly at z$\sim$6--8. This example is particularly revealing because of the flat (purple) SFH at $z=6$ from a CSF template that also produces \OIII and \Ha~emission, a burst that contributes the peak flux at $\lambda\sim1\micron$, and a $\sim$200~Myr-`old', aging burst from $z\sim8$ (green) burst template (but indistinguishable as a burst after $z=8$) that produces the rising flux beyond $\sim$3\micron. In addition to ``contiguous," the three epochs of star formation are close in mass at  2$\times10^8$\Msuns, a near constant level of SFH that might continue to $z=5$ and beyond.  We note that a small but significant fraction of these ``triples" could be fit with only two contiguous epochs of star formation, but none of them can be fit by a single epoch SFH. And, at the redshift most are seen, two or three epochs amount to 300-400 Myr in duration, long enough to host multiple generations of star formation.

This leads naturally into the the example in the lower right, where 4 epochs of the possible 6 have substantial star formation that is both continuous and, like the triple SFH, a coherent decline in added stellar mass.  The most important thing to recognize about these SFHs is that, again, they cannot be reproduced by the single or a pair of bursts of star formation like upper left and right, or by the shorter contiguous episodes that are shown in the bottom left panel (see also the Appendices B \& C of Paper I).   A long continuous SFH is defined here as (1) four epochs or more of star formation with less than factor-of-two uncertainty in stellar mass, and (2) free of two-epoch gaps.  The example shown here, and those that follow in Figures 2---5 are representative of 8\% of our $6<z<10$ sample.  It is important to reiterate that what matters for these longer, continuous SFHs is that they exist, rather than a dissection into gaps, spikes, dips, and wiggles. The data for this study is insufficient for \SEDz to deliver such detail, and indeed, this might not even be possible with higher resolution photometric filters or spectra, due to limitations in NNLS fitting.  For example, a single-epoch gap in one of the continuous histories is likely a consequence of ``non-negative solutions only": at any given redshift, adding star formation at that epoch may not improve the SED fit and might even degrade it. For this reason, the longer SFHs we find here are indicative of the general, not the detailed behaviour of star formation in the galaxy. 

The SEDs and solutions shown Figure 1 are exemplary: they display the very good fits of a majority of SEDs to single-epoch templates (bursts), and the ability of combinations templates to fit a range of SED shapes. That the fraction of cases that `fail to find a good solution' is small is, we think, strong evidence that the principle behind \SEDz is valid: an $A$-star dominated stellar populations can be used to recover accurate SFHs for galaxies in the $\tau < 1$ Gyr period of cosmic history. 

In the examples to follow, we will see that SFHs of significant duration are all found to be declining or nearly flat.  What is almost absent from the SFHs shown here is a stretch of \emph{rising} star formation. As we explained in Paper I, it is possible to attribute this to the difficulty of detecting a declining but `current' older population against a young one that was rising in flux but is now fading --- a simple selection affect. However, a more intriguing possibility is that the flat/declining phase of nearly all first-galaxy-SFHs --- from bursts to long SFHs --- were preceded by a strong, rapid burst of \emph{dusty} star formation lasting $\ls$50~Myr. Among the galaxies that were excluded from our sample we find 27 with steep, red SEDs, but \SEDz cannot be used to determine their redshifts, so their connection to the unobscured galaxies in our sample remains unknown.

\section{Choosing a Sample of 6 $<$ \MakeLowercase{z} $<$ 12 Galaxies}
\label{sec:Choosing a sample}

\emph{NIRCam's} exceptional performance and diversity of modes \citep{2023PASP..135b8001R}, have supported Guaranteed Time Observations (GTO) that cover wide variety of science programs, including exoplanets, star formation, and our Galaxy with its neighbors.  However, the largest component is devoted to the study of the early universe \citep{eisenstein2023overview}  --- the era of galaxy birth. The program described in this paper comes from deep-field imaging that is beginning to answer longstanding questions about how the first generation of stars --- collected into galaxies that served as reservoirs for the buildup of heavy chemical elements --- fundamentally changed the evolution of our universe.  

This study uses nine-band \emph{NIRCam} imaging of a $\sim$25-sq-arcmin field of the \emph{JADES }\emph{GOODS-S} survey. The images have been intensively processed, for calibration, to remove instrument signatures, combine dithered exposures, and characterize noise. These data were used to produced catalogs of objects (here v0.7.2) that identifies stars and galaxies, deblend overlapping images, and generates a number of different radial extractions of photometric measurements in 9 filters, \emph{NIRCam} bands F090W, F115W, F150W, F200W, F277W, F335M, F356W, F410M, and F444W. The catalog used here reached a depth of $\sim$30.0 AB mag and contained 24,350 extended sources.  Our study selected galaxies with $S/N>5\sigma$ (F200W \& F277W flux) within a 4-pixel-diameter circular aperture, appropriate for the small size of $z>5$ galaxies. Details of data quality, data reduction, and the production of photometric catalogs can be found in \citet{Tacchella+2023} and \citet{2023ApJS..269...16R}. To establish our sample, \SEDz was run on the complete \emph{GOODS-S} v0.7.2 catalog, with results sorted into four different redshift ranges, 6$<$z$<$7, 7$<$z$<$8, 8$<$z$<$10, and 10$<$z$<$13. To be precise, the actual ranges were shifted down by 0.25 in z, for example, $5.75<z<6.75$, in order to center on the epoch redshift (to match the templates), in this case $z=6$. {From the comparison of \SEDz redshifts with the ``known redshifts" of the synthetic \emph{Data Challenge}, SEDz was found to have an accuracy of $\Delta\sigma_z\approx0.15$ for $z>6$ galaxies.  This better-than-expected performance meant that interpolation between the templates used by \SEDz was justified and, in fact, not interpolating could add to systematic errors.  For this reason, boundaries were set as follows: z1: $5.75<z<6.75$, z2: $z<6.75<z<7.75$, z3: $7.75<z<9.75$, and z4: $9.75<z<12.75$.  Thus, when a galaxy's redshift was found to be within the range $5.75<z<6.25$, the z=6 template was used, but when $6.25<z<6.75$, a 50/50 interpolation between $z=6$ and $z=7$ templates was made (and so on for the other samples). In effect, this is equivalent to saying, the \chisq for $z=6$ and $z=7$ are the same (within errors), so the program averages the SED solutions (resulting SFHs), for those two redshifts.}\footnote{{In some cases, averaging the two solutions appears to have stretched out the SFH by one epoch, but we could not distinguish cases where this was the right thing to do, as opposed to shifting one or the other solution. This is likely to have produced some SFH3 types from 2-epoch SFH1 types. This potential and we believe uncorrectable error does not alter the outcome or conclusions of our study.}} {Eventually, data for the separate redshift ranges would be combined into a catalog spanning the full $5.75<z<12.75$ range, but the previous step allowed an investigation of performance over the redshift range that was helpful.}

By fitting combinations of stellar population templates, \SEDz found maximum-likelihood fits to SEDs that yielded redshifts in these four redshift intervals.  A well-recognized problem in deriving $z>4$ redshifts from SEDs is a degeneracy with $z\sim2$ galaxies where \emph{NIRCam}'s range of $1-5\micron$ translates into a rest frame coverage of $\sim0.3-1.7\micron$, typically covering a Balmer-break over an otherwise flat SED.  When very faint galaxies are the targets, these are easily mistaken for $z\sim6-8$ galaxies with a Lyman-break.  We used three different methods to mitigate the problem.  First, we wrote code in \SEDz that compared the shape of the \chisq curve when \emph{two} minima were found, one for $z\sim2$ and another for the higher redshift. During the `Data Challenge' tests, we found that using the overall slope of the run of \chisq with redshift, the depth and width of the minima, and the color of the \SEDz itself, removed roughly 50\% of cases. We also used archived \emph{HST} data to find at least two 2.5$\sigma$ detections in three  visible bands of the \emph{WFC3} imaging, F606W, F775W, and F814W.  Flux below the Lyman-break is the best rejection method, but we have found that, for galaxies this faint in the near-infrared, only about one-third of low-redshift galaxies are detected in these bands, even with the deepest \emph{HST} imaging available.  The third check was to use EAZY `photo-z' redshifts for the cataloged objects, part of \emph{JADES} team data processing for internal and eventual community use \citep{Hainline_2020}.

{The procedure was to run \SEDz for each redshift interval using only (1) the rejection of low-z objects by detection of visible flux and (2) the internal \SEDz $z<4$ rejection.  This produced four subsets with 759, 374, 277, 61 galaxies at z6-7, z7-8, z8-10, and z10-13, defined in Section~\ref{sec:Choosing a sample} as z1, z2, z3, and z4 (shifted by -0.25 from integer z values). Taking the \SEDz redshift (hereafter, $z_{SED}$) as the adopted redshift was required, because that is the value for which the SFH is derived, but comparing \zSED with EAZY-derived redshift, \EZ, and finding them consistent was taken as the next level of ``qualification."  This was termed the ``gold sample," and amounted to 446, 183, 115, and 15 `confirmed' objects."  This left 313, 191, 162, and 46 objects with \zSED \emph{unconfirmed} by \EZ. Their SEDs were inspected, one-by-one, to decide if the Lyman-break range was sufficiently well defined to suggest a low-redshift for the galaxy.  If so, the 7 wide-band images (readily accessed through a `\emph{FitsMap}' viewer from the \emph{JADES} image-processing team) in particular, the F090W, F115W, and F150W images, were inspected in order to evaluate \emph{visual} evidence for a Lyman-break. From these $\sim$712 inspections, 225, 123, 103, and 38 galaxies were rejected. The remaining 88, 68, 59, and 8 galaxies were added to the ``gold sample," based on judgement that at least half of these were at the higher redshift found by \emph{SEDz*}.  The count for the four redshift slices was then: z1---534, z2---251, z3---174, and z4---23, a total of 982 galaxies.}

{Subsequent to the first release of this paper, we ``pruned'' the data sample further.  An additional 72 galaxies were deleted from z1, 18 were deleted from z2, and 2 were moved from z2 to z3.  About half had been removed in ‘FitsMap’ inspection for proximity to diffraction spikes; the remaining half was split between marginal `S/N$\approx$5' cases and poor overall fits of the model to the data. The final numbers for this study are z1---462, z2---233, z3---176, and z4---23 (unchanged), a total of 894 galaxies.

\section{Further Examples of SFHs} 
\label{sec:Examples SFHs}

To expand on the introduction of the four SFH types we showed in Figure~\ref{fig:4SFHs}, we show 18 examples for each of four redshift ranges in Figures 2, 3, 4, and 5. In each, there is a row-by-row progression, starting from the top, from a single burst population, through stochastic (multiple bursts), to contiguous epochs of star formation, and finally longer, continuous over four or more epochs.

{\emph{SEDz*} plots, like the ones that follow, and their associated data, will be available for the full 894-galaxy sample at \url{https://obs.carnegiescience.edu/SEDz-star/SFHs}}.

{Beginning with Figure~\ref{fig:SFHs-z6}, we see three examples of a $z=6$ starburst, each with a mass of $\sim$2$\times10^8$\Msuns.}  {As in the discussion for the burst SFH in Figure~\ref{fig:4SFHs}, a single epoch of star formation --- both occurring and observed at $z\sim6$ --- is a very good fit to each observed SED\footnote{Recall that the magenta band shows the quartile ranges of solutions while the single purple line shows a \emph{single} solution that is used to break down the SED into its component parts, epoch by epoch.}. For all three, the solution is a combination of a burst and continuing ``constant" star formation (CSF) with less than a 50\% contribution to the flux(denoted by the small green arrow).  (A confirmation of the redshift found by \SEDz is the \OIII emission in the top left case, detected in the medium band filter F335M.) Again, the excellent fit found for such cases validates that the stellar population templates are correct for the task.  Single bursts are the most common SFHs in our study --- 473 cases, 53\% of the full sample.}

{We also identify the next two examples (second row, left and center, hereafter, `2l, 2c') as \emph{single bursts}.  Their \emph{two-adjacent epochs of star formation}, may be unresolved, that is, one event that is best fit by using consecutive templates.  In both 2l \& 2c the burst at $z=7$ would have been very blue, but one epoch later -- when these galaxies are observed --- their contribution rises to the red, causing the elevation in each spectra.  It seems reasonable that star formation episode of $\tau\sim100$ Myr could produce this \SEDz result by spanning the relatively long ``integer-redshift" epochs of $z=6$ and $z=7$.  Under this interpretation of a single event, we include later examples where the two masses differ by as much as an-order-of magnitude, implying a sharply rising or falling episode of star formation.}

\begin{figure*}
\centerline{
\includegraphics[width=7.3in, angle=0]{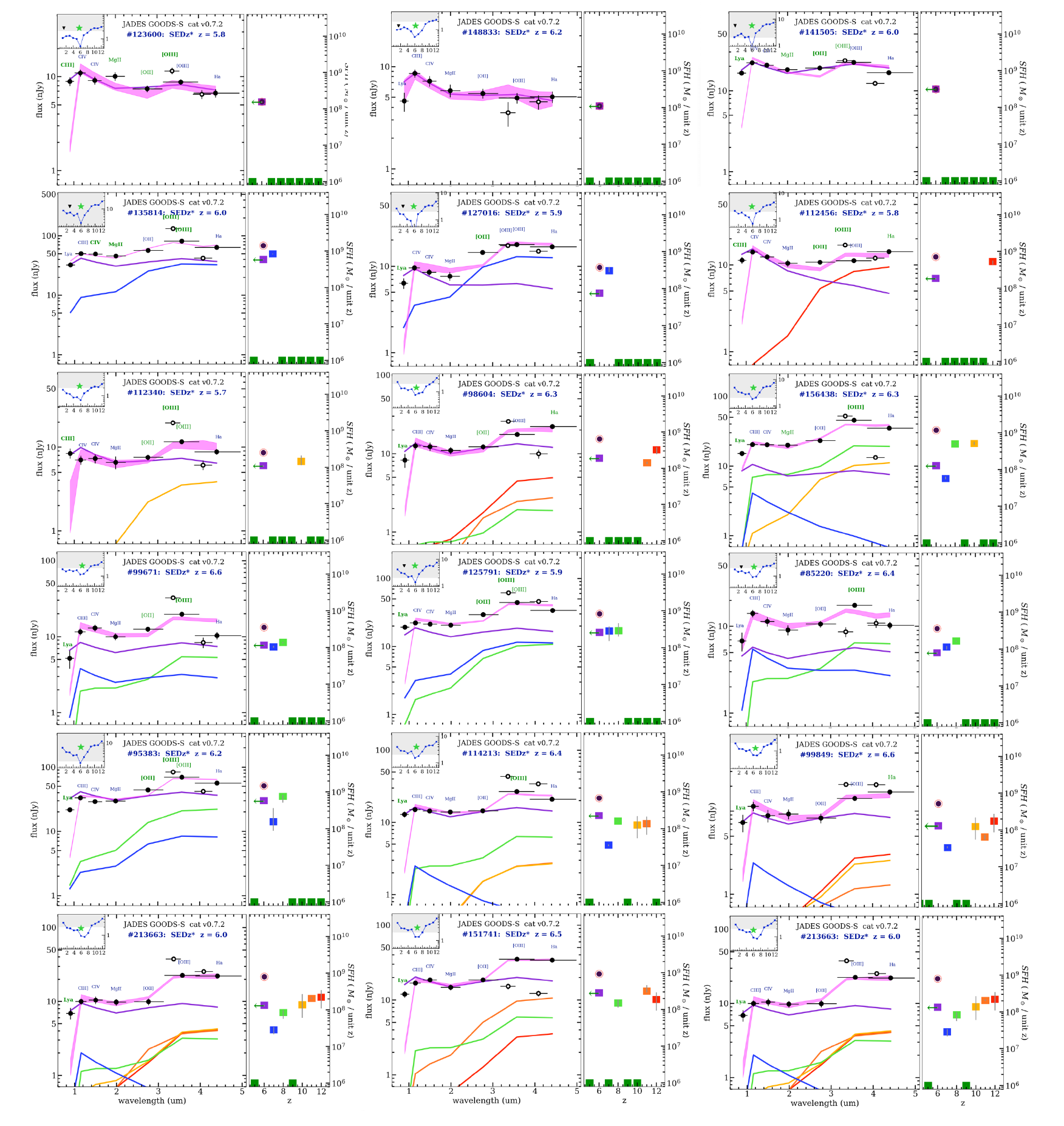}
}
\caption{Examples of the identified four types of SFHs of this study observed at redshifts $z\sim6$ ($5.75<z<6.75$), starting with three examples of single starbursts in the top row and continuing in order, across the rows and and down, to multi-burst ``stochastic" histories, three-epoch \emph{contiguous} runs, and finishing with long SFHs of four epochs or more.  Detailed explanations of the salient characteristics of these types are described in the text.
}
\label{fig:SFHs-z6}
\end{figure*}

\begin{figure*}
\centerline{
\includegraphics[width=7.3in, angle=0]{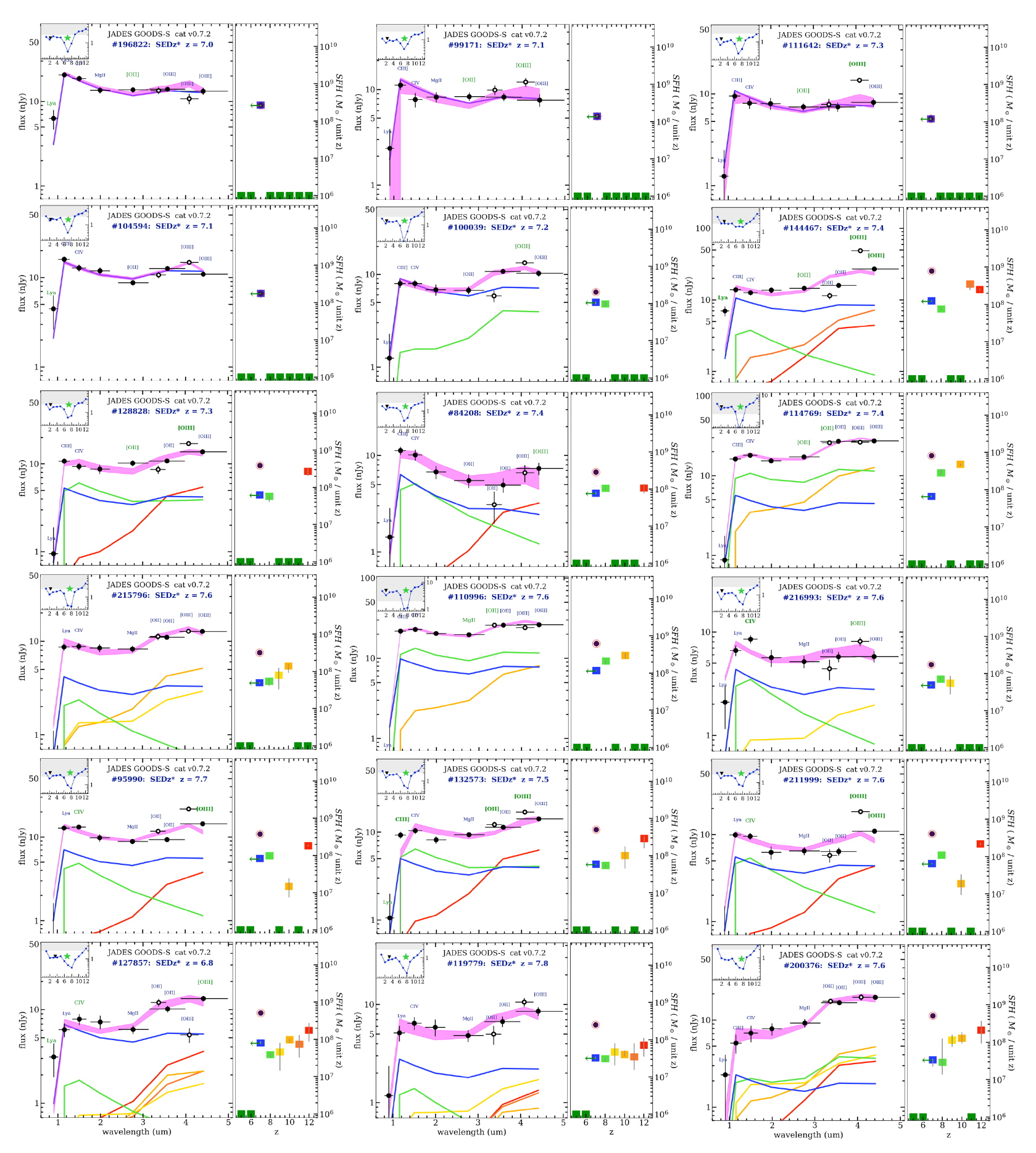}
}
\caption{A virtual `replay' of the types in Figure~\ref{fig:SFHs-z7} for 18 galaxies at $z\sim7$ ($6.76<z<7.75$), as described in the text. This set includes five examples with very high S/N [Row(1-5):position(left-right-center)] = 1l, 2l, 2r, 3r, and 4c, that are exquisitely fit by \SEDz with its ``present-epoch" templates, demonstrating the fidelity of the SFHs that \SEDz can deliver.
}
\label{fig:SFHs-z7}
\end{figure*}

{The next examples, which exhibit well separated bursts, are additional demonstrations of the ability of \SEDz to reveal dominant early star formation in galaxies observed $\sim$500 Myr later, giving confidence in its identification of SFHs that extend over the $6<z<12$ range of our study. Comparing the stellar population templates shown in the \emph{Appendix B}, it is clear that these three examples show histories that cannot be reproduced by \emph{any single stellar population}.  Figure~\ref{fig:SFHs-z6} shows 3 examples (2r, 3l, 3c) of  strong, well-separated bursts of star formation.  We call these \emph{stochastic} SFHs because it seems unlikely that these separate bursts can be ``tied together,"  not even if there was weaker, undetected star formation in-between (which would be at least an order-of-magnitude smaller in mass). The following SFH (3r) is interpreted as either 3 or 4 epochs of star formation starting at $z=10$).  All four are graphic illustrations of very young, blue stellar population whose red flux has been `boosted' by older populations, at $z=10-12$.  For all four of these \emph{stochastic} SFHs, redshifts are again confirmed by \OIII emission in the F335M band.  Stochastic star formation, by our definition, is almost as common as the single bursts (in total epochs of star formation), adding 165 cases (19\%) of two or more bursts.  Together, `bursts' and `stochastic' make up 72\% of SFHs found here: clearly, this is a strong and, we believe, unexpected, \emph{dominant mode} of galaxy building in its beginnings.}

{However, the bottom 3 rows of Figure~\ref{fig:SFHs-z6} remind us that a significant fraction of early galaxies are undergoing more orderly, gradual growth.  In examples 4l, 4c, 4r and 5l three stellar populations are required to fit the observed SEDs. SFH3 star formation is usually ``contiguous," but sometimes with a one-epoch gap, likely due to `noise' or the asymmetry in NNLS solutions --- no negative star formation. (See Figure 3, the gap in 4c, but note the $z=9$ star formation 4l that shows the opposite --- a \emph{detection} that is dubious). SFH3 describes 21\% of full sample (25\% of the $z=6$ sample). All three have final star formation at $z=6$ with and added CSF components, hence the very flat purple CSF contributions (see \emph{Appendix B}). Again, it is also clear that no amount of $z=6$ or $z=7$ star formation can produce the `rising-to-the-red,' in all but 4r, with its uncertain star formation at $z=8$  and noisy SED.  Although they look short, these last 3 epochs add up to half-a-billion years, or half of the time since the Big Bang, so these are a substantial departure from what seem to be the dominant SFH mode --- $\ls$100 Myr starbursts.} 

{The final 5 examples, 5c, 5r and row 6, are longer, more-continuous histories (SFH4) that stretch over the full redshift range, with 4 to 6 epochs of star formation. Although two have a 2-epoch gap that could have landed them in the `stochastic' category, for all there is an orderly history of constant or declining star formation.  These are pronounced examples of young stellar populations with a history of very early star formation that accounts for their strongly `rising-to-the-red' SEDs. All five show $z=11$ and $z=12$ star formation, however it is likely that this period of $\sim$100 Myr is not ``resolved,'' so the SED could be reproduced by only one, with the combined mass.  Although they represent only 3\% of the $z=6$ population and only 69 galaxies in the 894-galaxy sample (8\%), the persistence of SFH4 over the full time range explored here suggests a different environment, one where galaxies can evolve more slowly and relatively undisturbed, for example, experiencing only minor mergers.}

From this first set we see the prevalent signature of early (z $\gs10$) star formation: a SED rising to the red, often accompanied by substantial star formation at OE.  This is best seen by the different levels for the medium bands F335M and F410M than is defined by the 7 broad-bands.   This raises the question of whether strong emission lines could be boosting the far red fluxes and mimicking early star formation. We show in \emph{Appendix C} that this should not be the case, because the CSF templates \emph{include} active star formation, so that the level of the broad bands cannot be `raised' by emission --- it's already included.\footnote{The caveat is emission is much stronger than in the present-epoch star-formation \SEDz templates: the ``rise-to-the-red" could be mistaken for star-formation hundreds of Myr earlier.  This effect is likely to be seen in the  present study, at a level of `factor-of-2 or less' excess stellar mass for a $<$10\% fraction of the SFHs (see \emph{Appendix C}). For example, SEDs 5l and 5r of Figure~\ref{fig:SFHs-z7} are very similar, with strong \OIII\ emission the only high points, suggesting that the SED rise is due to stronger emission than in the templates.  In contrast, the center SED  has a steadily rising fit that is certainly the result of the $z=12$ star formation.}  The medium bands F335M and F410M (open circles --- not used in the NNLS fit) provide good reference points for the presence of emission and the level in the continuum, because emission-line strengths were not calculated for the narrower bands. \emph{Appendix C} provides more examples and discussion.

For further examples of SFHs, Figure~\ref{fig:SFHs-z7} shows for $z\sim7$ a virtual replay of the $z\sim6$ population: 5 individual bursts (4 single, 1 `twin'), 4 well separated ``stochastic" events, 3 more triples, 3 noisy SFH3/SFH4 histories, and 3 long SFHs, none with a gap of more than 1 epoch, and two cases with star formation in all 6 epochs. (Some have large error-bars that remind us that these are ``representative" star formation histories, not to be considered faithful ``epoch-by-epoch.)"  As in Figure~\ref{fig:SFHs-z6}, but not explicitly called out, are very high S/N examples, in particular, examples 1l, 2l, 2r, 3r, and 4c are well-defined SED solutions that match the fluxes of the data with remarkable fidelity.  Considering the relatively crude redshift-resolution of the stellar population templates, the agreement of model and data is convincing evidence that the modeling works and the templates are fully ``descriptive:" 3 or 4 templates are usually all that's needed, and that's good, because that's all that are available!

\begin{figure*}
\centerline{
\includegraphics[width=7.3in, angle=0]{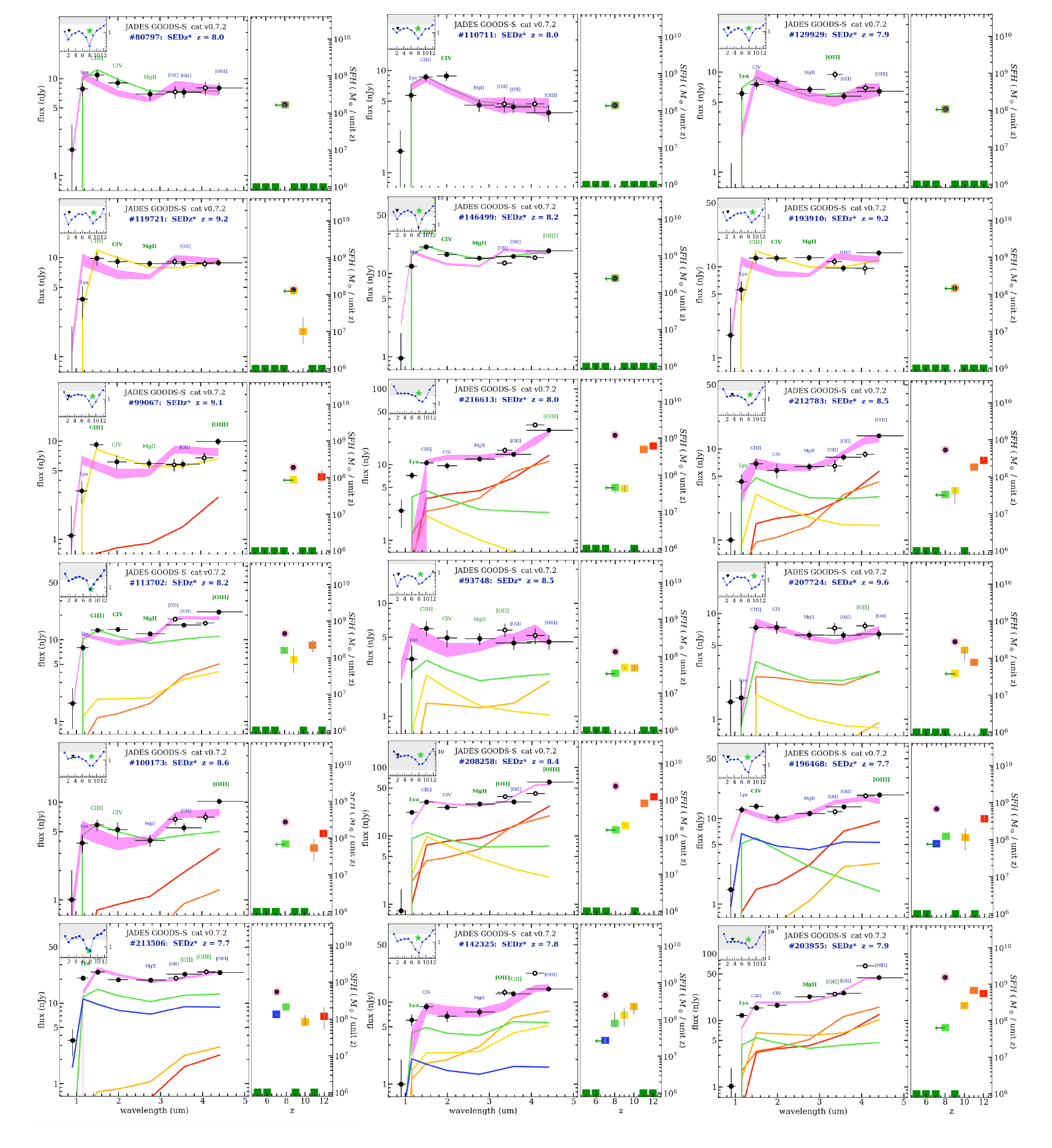}
}
\caption{The persistence of the four SFH types reaches back $\sim$700 Myr from $z\sim6$.  No obvious \emph{evolution} of the mix of types is apparent, although changes in ``proportions" may be appearing (see Figure~\ref{fig:SFHs_across_time}).  At this earlier epoch the prominence of the Lyman-break is a strong factor in finding the redshift and establishing the SFH.  More than half of these $8<z<10$ galaxies show detected star formation back to $z\sim11-12$.
}
\label{fig:SFHs-z8-10}
\end{figure*}

\begin{figure*}
\centerline{
\includegraphics[width=7.3in, angle=0]{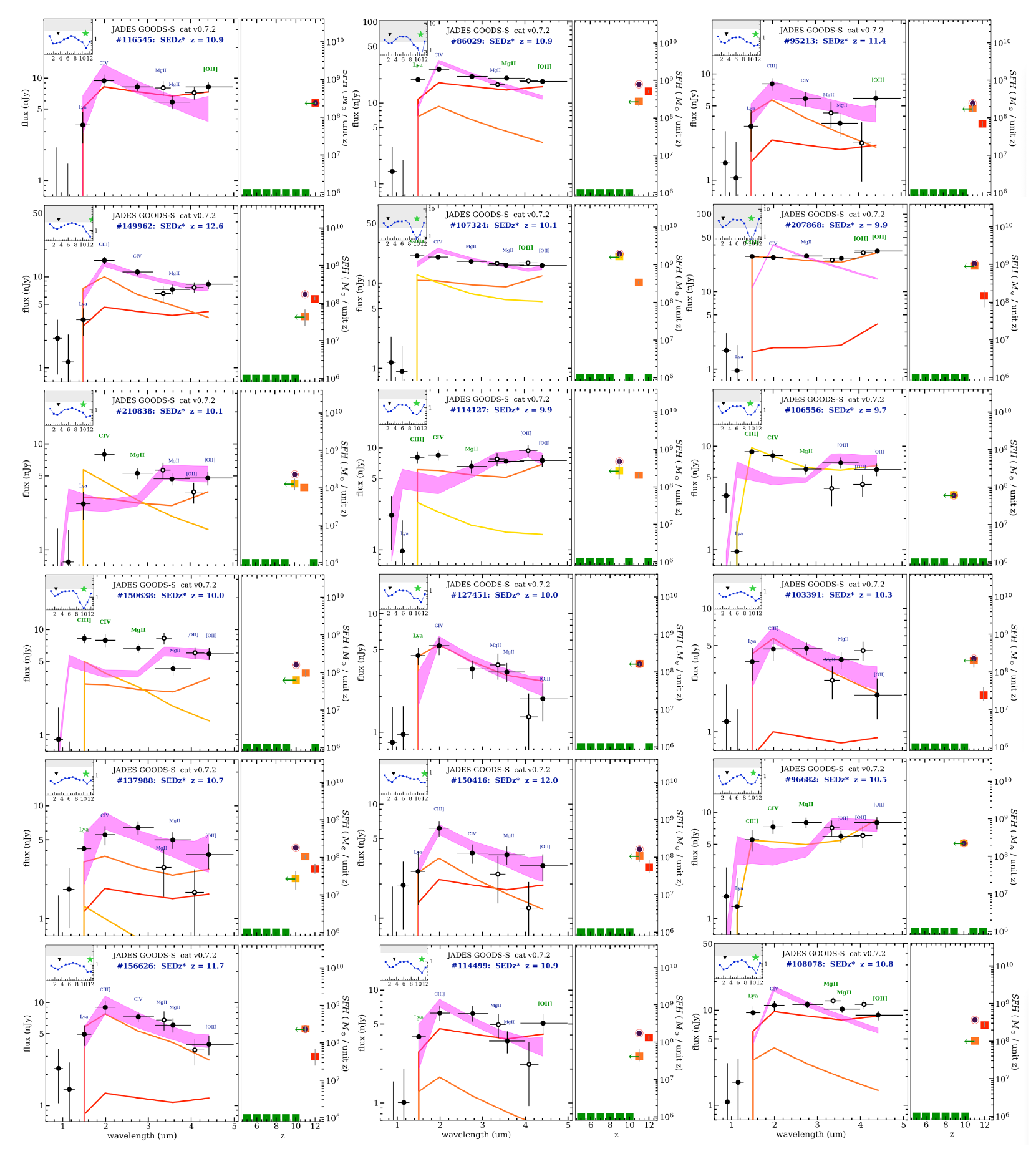}
}
\caption{18 of the 25 examples $z\sim10-12$ in our 894-galaxy sample.  The Lyman-break now extends from F090W into the F115W filter as well.  The timescale covered in this figure is only $\sim$200 Myr, shorter than most of the longer histories we find at lower redshift, but there are three cases --- 2c, 3c, 5l --- that seem to cover this whole period.
}
\label{fig:SFHs-z10-12}
\end{figure*}

Extending the redshift coverage up to $z=8-10$ in Figure~\ref{fig:SFHs-z8-10} shows that the classification into the four types of SFHs continues to earlier times.  All are represented in similar frequency to those at lower observed redshift. The Lyman-break is, of course, pronounced, which helps with constraining the \SEDz fiting. Most of the the bursts (single and double) in the top three rows rise sharply to the (rest-frame) ultraviolet --- signaling very young populations, but conversely, the bottom two roles show the steepest rise to the red in the sample, in this case signaling strongly declining rates of star formation since $z=12$.  

Finally, in Figure~\ref{fig:SFHs-z10-12}, we see 18 of the 23-galaxy-sample of the highest redshifts.  The Lyman-breaks are clifflike, and the SEDs are all blue within this sample that covers only $150$ Myr, only $\sim$200 Myr since $z=20$ when the first ``modern" galaxies were likely born.  Rapid evolution of the universe at this time, especially the strong growth of dark-matter halos and rapidly decreasing density of large-scale structure, suggests star formation that might be itself changing in character or composition.  Yet, remarkably, the stellar populations we observe are all matched by the stars that we have all around us today, some 13 billion years later.

\section{SFHs Across Time}
\label{sec:SFHs_Across_Time}

Does the population of SFHs itself `evolve'? Figure~\ref{fig:SFHs_across_time} plots SFH type for 894 different galaxies as measured at their epoch of observation, color-coded to show burst SFHs as red and orange (single and multiple), with contiguous `triples,' and continuous `long SFHs,' as green and blue.  To first order, we see two things: (1) the ``final" masses of all 894 in our SFHs sample are largely confined to $10^8$--$10^{10}$\Msuns, and (2) all four types are represented over the full redshift range covered in this study.  That there are some ``accumulated" stellar mass below $10^8$\Msuns, but none below $5\times10^7$\Msuns, is a simple detection limit: the fluxes would generally fall below the $S/N > 5$ limit we have chosen.  On the other hand, the steep falloff in numbers above $M>3\times10^9$ \Msun is probably a reflection of  astrophysics, for example, stellar feedback suppression in the environment of rapid star formation in the compressed volumes of these sub-kpc-sized sources.  Powerful feedback is probably expected in the case of the bursts, but it is less obvious why the longer histories, with lifetimes of hundreds of Myr, would be subject to the same limitations, and yet the distribution here suggests, interestingly, that they are.  This, and the clear result that most of the stellar mass in this formative time is made in relatively large starbursts, should place strong constraints on numerical simulations of galaxy growth.

The `evolution' of the proportions SFHs over time is harder to parameterize, but it is clear from the fact that all types show up over the diagram that there is no strong evolution of the \emph{population} of SFHs.  Figure~\ref{fig:SFHs_rough_dist} shows a crude graph of the frequency of the four SFH types for the sample in Figure~\ref{fig:SFHs_across_time}. With the exception of the 70\% fraction of bursts at the highest redshifts\footnote{This probably not a result of small-number statistics as  it is the smaller interval of time to express the four SFH types}, the variation in the rest of the plot is `factors-of-two' and not much in the way of trends.

We should add that there is some non-statistical ``bunchiness" in Figure~\ref{fig:SFHs_across_time}, such as the `swells' of SFH3 `contiguous' at $z\sim6.5$ and SFH4 `long-SFHs' at $z\sim7.5$, and the `valley' of single-bursts at $z\approx6.5$, but we believe that these are more likely \SEDz ``preferences" associated with interpolating between the epochs, rather than a real effect.  A denser sampling of epochs, with twice the time resolution for the templates, would likely resolve this, but for now we focus on broad trends.

The lack of strong change over time (for this early period) provides little in the way of clues about the nature or ``causes" of the different histories.  This suggests that dependencies based on space, rather than time, might provide more insight.  We investigate this in Section~\ref{sec:SFHS Space}.

\begin{figure*}
\centerline{
\includegraphics[width=6.3in, angle=0]{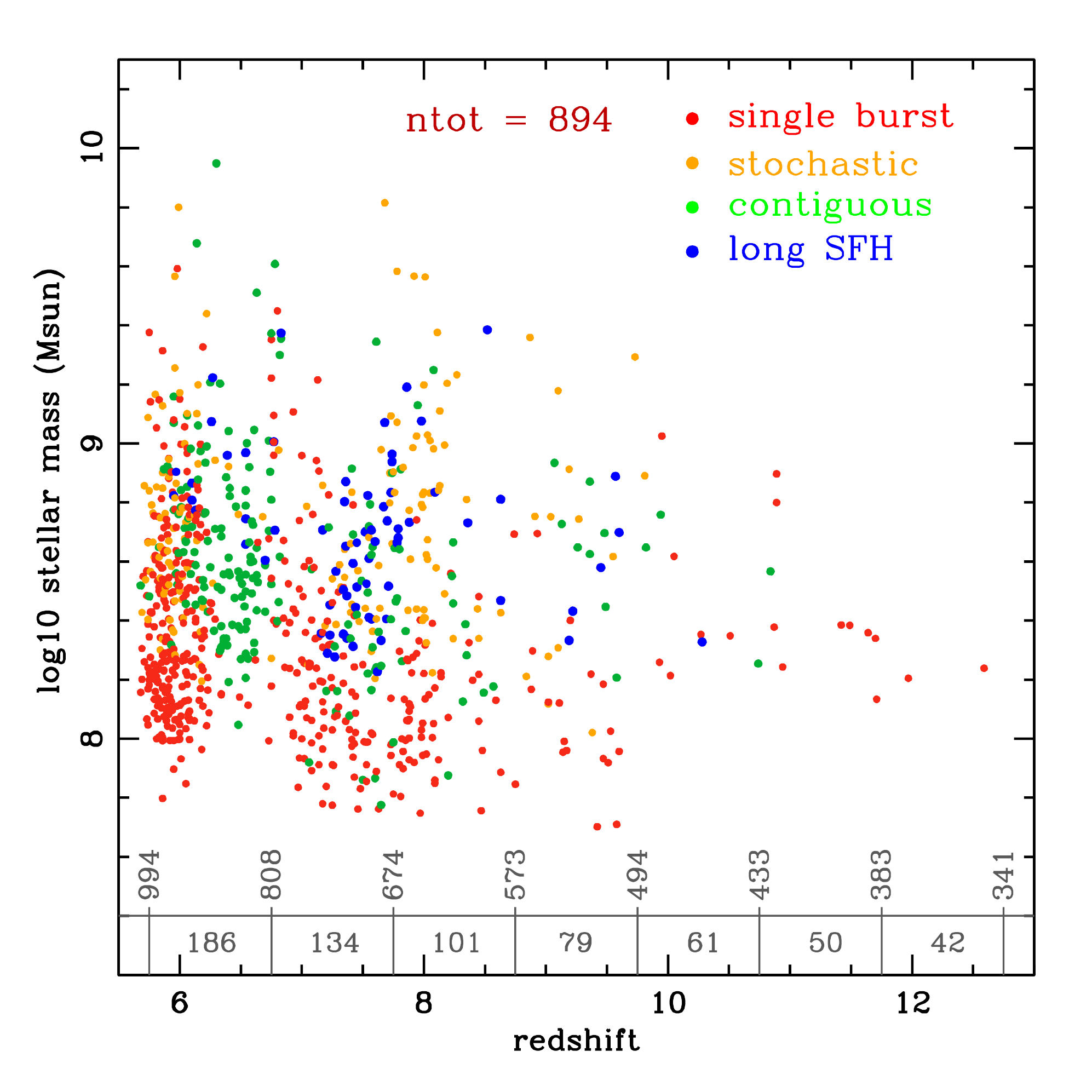}
}

\caption{The stellar masses associated with the four SFH types displayed over the full $6<z<12$ time frame of this study. The bottom scales are the age of the universe in megayears, and the duration of each of the 7 epochs of this study.  Each dot represents a galaxy at the epoch-of-observation, at its final mass, with SFH type indicated by dot-color. (Figure~\ref{fig:mass_buildup_all} \& Figure~\ref{fig:mass_buildup_all_integrated} show the period spanned \emph{by each SFH} by ``connecting the dots" from the beginning to end of stellar mass growth.) The most important feature in this figure is the presence of all four types over the full range of epochs.  Proportions of the four SFH types appear to change over the distribution (for a rough picture, see Figure~\ref{fig:SFHs_rough_dist}), like the apparent ``swells" in long-SFHs (SFH4) that appear at $z\sim7.5$, or contiguous histories (SFH3) at $z\sim6.5$.  These may be the result of a biases in \emph{SEDz*} or, for example, large-scale structure, but the full coverage of each SFH type in both epoch and mass is the takeaway.
} 
\label{fig:SFHs_across_time}
\end{figure*}

\begin{figure}
\centerline{
\includegraphics[width=3.3in, angle=0]
{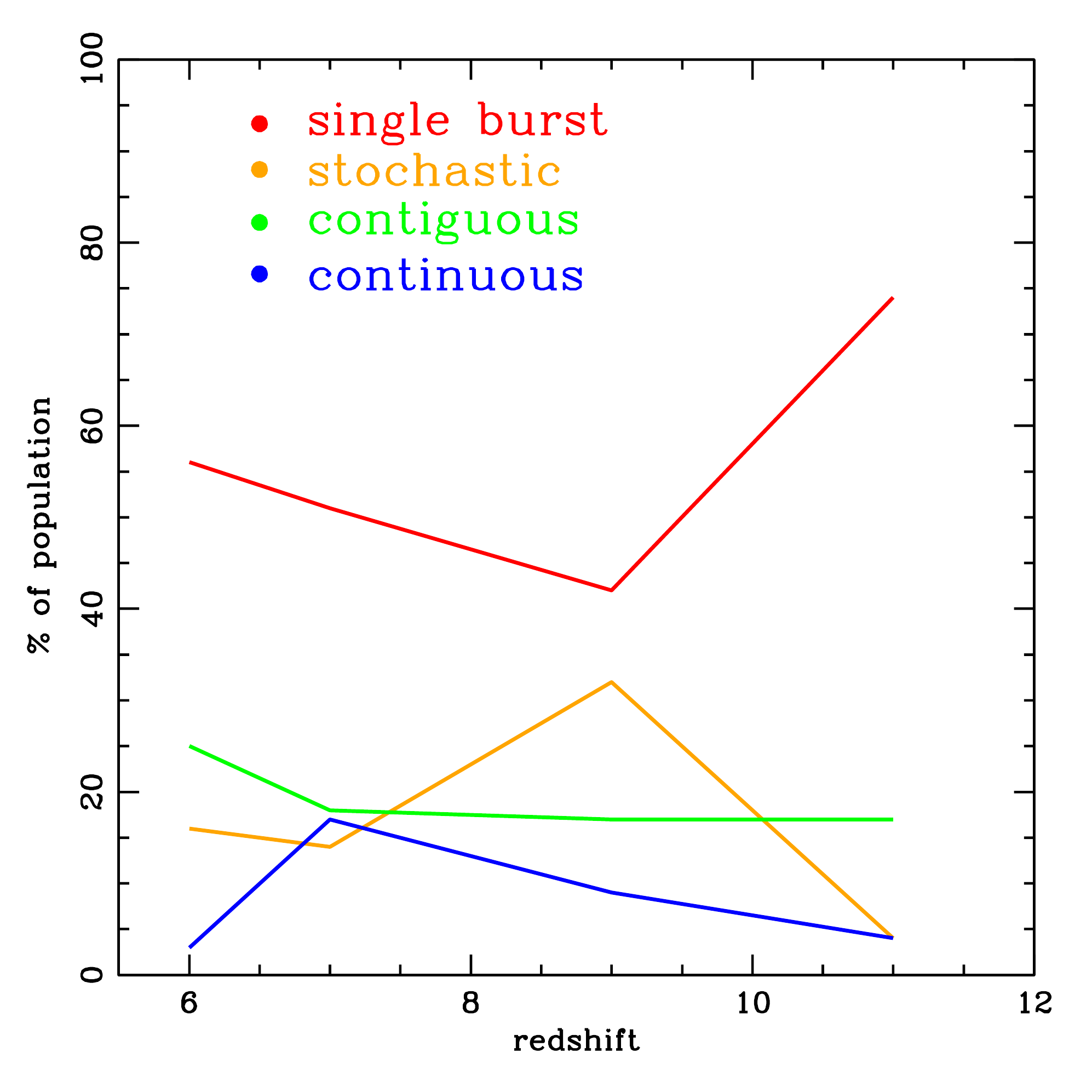}
}
\caption{The rough proportions of the four SFH types with redshift, derived from the distribution in Figure~\ref{fig:SFHs_across_time}.
} 
\label{fig:SFHs_rough_dist}
\end{figure}

\begin{figure*}
\centerline{
\includegraphics[width=7.0in, angle=0]{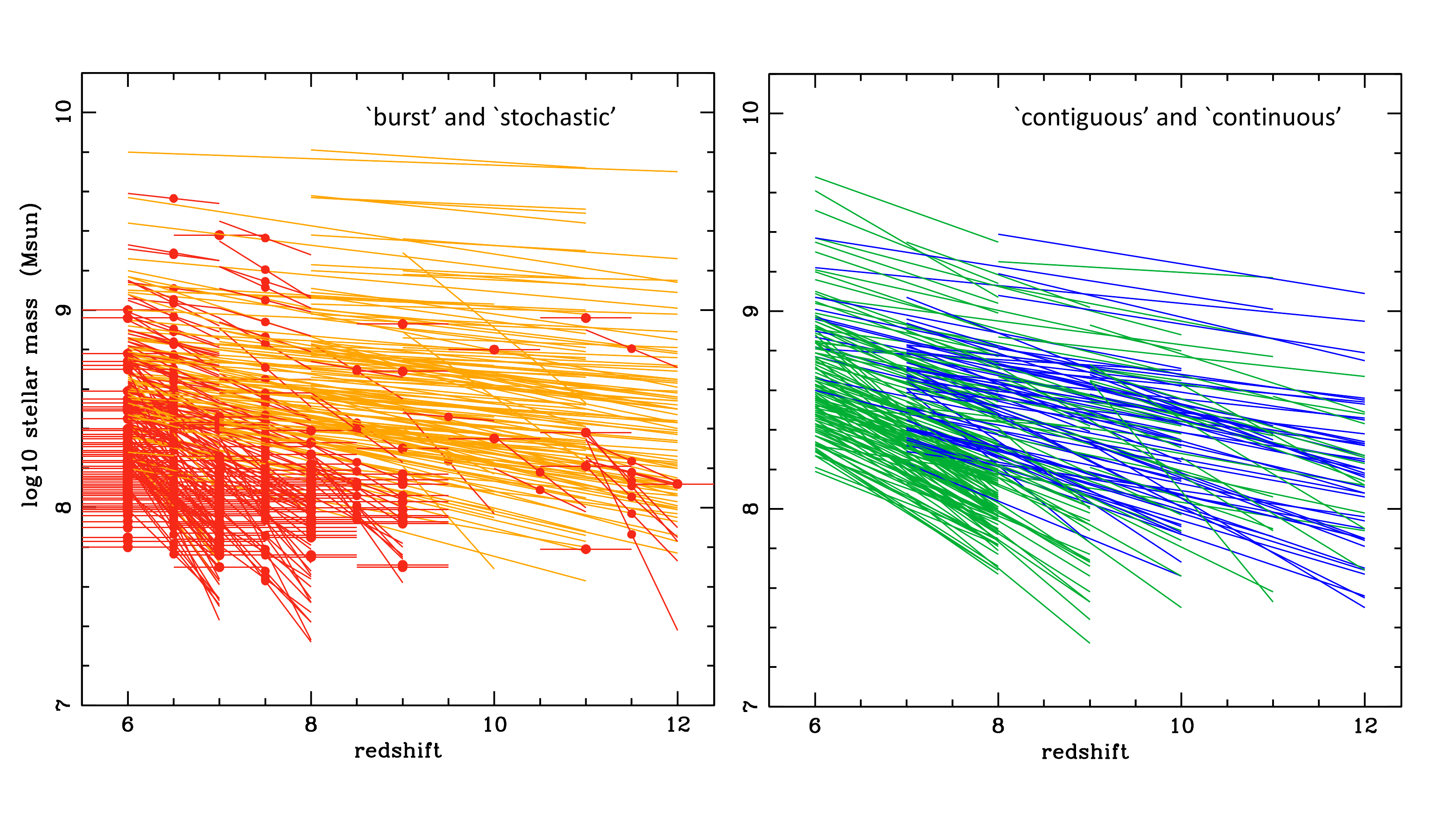}
}
\caption{(\emph{left}) The contribution to the stellar mass during the period $6<z<12$ from galaxies dominated by bursts. Galaxies with SFH1 histories are red dots; with `handles' mark their epoch. Orange tracks --- SFH2 --- begin with the first epoch of star formation and connect to the last, showing that such systems do not grow more than factors of 3 or 4.  Instead, new systems appear: that is how total stellar mass grows during this period. The single burst cases do not add substantial mass until $z\sim9$ (see Figure~\ref{fig:mass_buildup_all_integrated}.) (\emph{right}) The longer SFHs, both 3-epoch contiguous (SFH3) and 4-or-more-epochs continuous (SFH4) show substantial growth from $z\sim12$ to $z\sim6$, reaching a similar level of contributed mass in the range of $10^8$--$10^9$\Msun (Figure~\ref{fig:mass_buildup_all_integrated}), but a factor of 3 less than the SFH2 multiple bursts.  For both SFH3 and SFH4, the stellar mass accrues over $\sim$0.5--1.0 Gyr, perhaps suggesting less volatile surroundings and circumstances in their development compared to the SFH1 and SFH2 burst histories.
}
\label{fig:mass_buildup_all}
\end{figure*}

\begin{figure*}
\centerline{
\includegraphics[width=6.0in, angle=0]{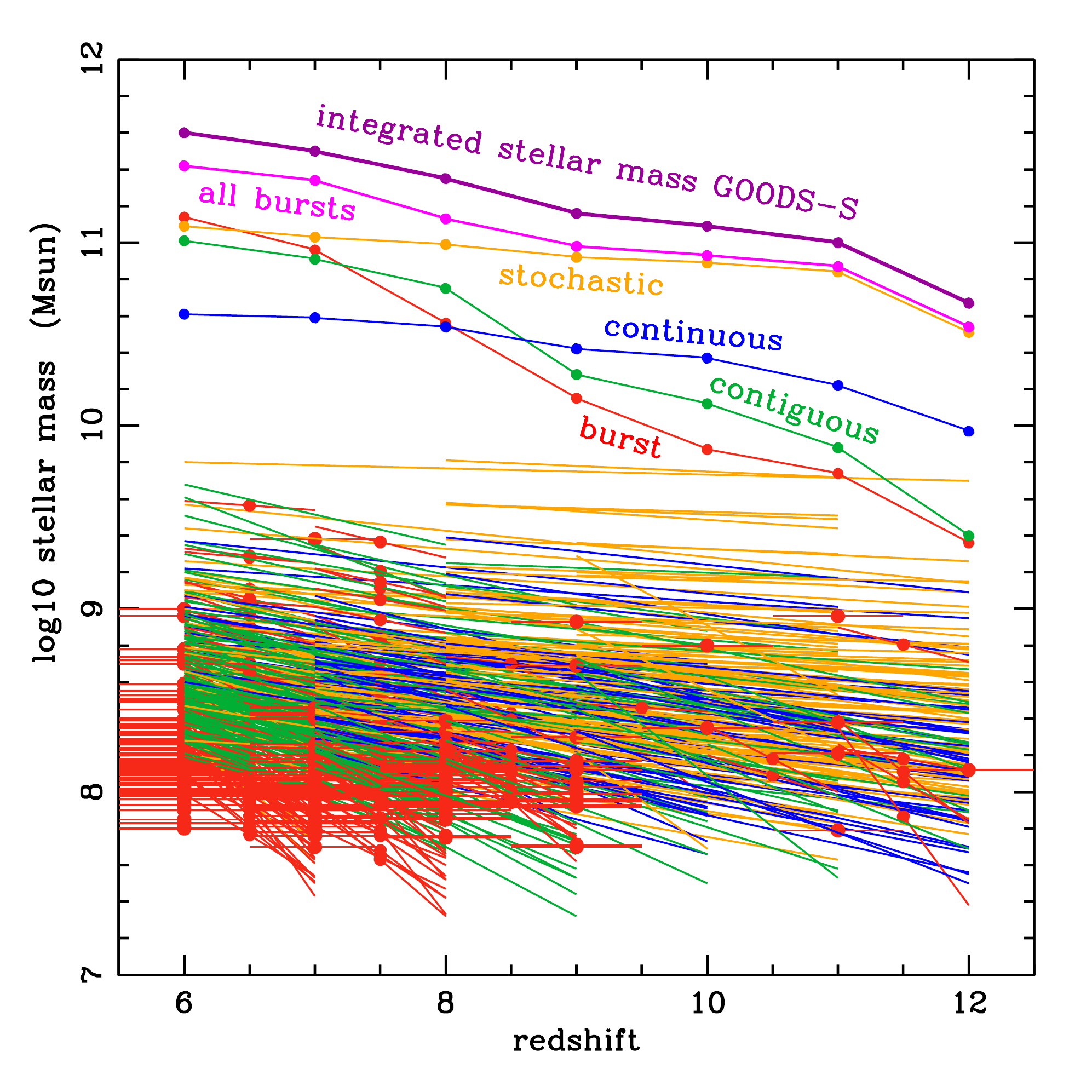}
}
\caption{The combined diagram showing all (beginning-to-end) histories, summed  to produce the integrated mass (recorded as the colored lines above). The substantial growth in stellar mass that happens in this volume of space from $z=12$ to $z=6$ is mostly from adding sources, amounting to the 894 we collected for our study.  SFH1 and SFH3 rise rapidly and reach nearly the same value of $\sim$10$^{11}$\Msun. SFH2 accumulates as much mass, but is flatter because, typically, these begin with an strong early burst and end with a much-later, smaller one.   The contribution of SFH4 is also flatter, rising by only a factor of $\sim$3, due to declining mass contributions, in this case.  SFH4 histories contribute a factor-of-three less mass than the others, the result of their lower frequency of 8\%.  Making the principle point of this paper again, SFH1 \& SFH2 combine to contribute 2.6$\times$ 10$^(11)$ \Msun of the total stellar mass --- 66\%.  Starbursts rule.
}
\label{fig:mass_buildup_all_integrated}
\end{figure*}

\section{A Galaxy is Born}
\label{sec:Galaxy born}

Returning to our theme, we want to explore how the results obtained in this study can help in understanding how galaxies began to grow and build up the essential elements for planets and life.  Specifically, how has the progress of star formation proceeded in the first billion years of cosmic history, birthing and growing new galaxies in a environment more gas-rich and more turbulent than we easily imagine?  At the same time, these young galaxies are likely fed continuously by smooth inflow of gas, incorporating less massive galaxies and adding their moderately metal-enriched stellar populations.  All this continues as young galaxies are vulnerable to violent major mergers and the huge energy released in massive starbursts, and possible large-scale changes through black hole formation and growth.  It is not likely that such questions will be answered solely through observations but, rather --- as always --- theory will be required to explore the physics of each of these elements.  Numerical simulations should benefit greatly from these kind of data, replacing previous and various speculations, where the manifestations of star formation that set the course are reproduced, and, we hope \emph{understood}.

What we have found in this study already confirms the dynamism of the epoch where galaxies achieved masses of $10^8$--$10^{10}$\Msuns. We find large contributions to the growth of stellar populations by bursts unlike any we see today, strong enough to produce a stellar mass of $10^8$--$10^9$\Msun in an episode lasting only $\ls$100 Myr --- little more than a dynamical time --- and, apparently, strong enough to ward off further star formation for more than $\sim$500 Myr, and maybe even a Gyr. And yet, we also see common cases of multiple bursts over which the total stellar mass can reach well over $10^{10}$\Msuns.  What is the difference, then, between these and systems making the similar amounts of stellar mass, but over the same long period of time where bursting galaxies seems to go dormant?

A good way to appreciate the power of these data can be seen in Figure~\ref{fig:mass_buildup_all} and Figure~\ref{fig:mass_buildup_all_integrated}, where we plot the mass buildup over $6<z<12$ from these different modes of star formation. These plots use the \SEDz star formation histories for our 894-galaxy sample to graph the onset of star formation and, in the cases of longer SFHs, its subsequent addition of stellar mass.  Here we have ``connected the dots'' --- from when star formation began to the last epoch where it is detected --- to show stellar mass buildup. (Color coding is the same as for previous figures.)

Figure~\ref{fig:mass_buildup_all} (left) shows this for single bursts and bursty `stochastic' histories, SFH1 \& SFH2, and (right) longer ``contiguous" and ``continuous" histories, SFH3 \& SFH4.  In an effort to provide guidance to numerical-simulation modelers trying to answer such questions, we now express the data we have described here in terms of the growth of galaxies with such different histories in mass-buildup diagrams. The plot is for mass versus time, as in Figure~\ref{fig:SFHs_across_time}, but now with tracks that connect the first and last epochs of recorded star formation within the $6<z<12$ era.  In Figure~\ref{fig:mass_buildup_all}, SFH1 --- a single burst, shows up as a dot with a handle marking its epoch, while an SFH2 --- ``stochastic," in orange, appears as a shallow rise, since the two bursts are usually well separated.  It is also easy to see that, while the single bursts are the most common SFH type, multiple bursts are both larger in mass to begin with (we find mostly declining SFHs) growing with subsequent bursts, such that they add the most mass over most of this era.  A critical point, though, is that the mass from these burst-dominant galaxies is growing through the appearance of new bursts.  The most frequent number of bursts for SFH2 galaxies is just 2 --- a single added burst rather than a several. Stellar mass is primarily growing by adding bursting objects, single and multiple, not by many smaller bursts in each galaxy.\footnote{Recall that, although we use a burst to represent each epoch of recorded star formation, the stellar mass formed is -- from later epochs -- unresolved.  The ``dot-like" representation of SFH in the \SEDz plots is only symbolic.} The same diagram for the 3-epoch `contiguous' (SFH3) and 4-epoch or more `continuous' (SFH4) histories shows more tilt in its tracks: this is most obvious in the 3-epoch tracks that become more and more dominant from redshift $z=9$ down.

Putting it all together in Figure~\ref{fig:mass_buildup_all_integrated}, we combine the burst and longer histories, summing and integrating to learn how the stellar mass of this collection of 894 galaxies grew from $z=12$ to $z=6$. We see the contribution of single bursts grow strongly, while the average mass for SFH1 --- the lowest of the 4 types --- does not increase over the full redshift range.  By $z\sim6$ the single-burst, stochastic, and continuous histories have each added $\sim$$10^{11}$\Msun to this volume of space.  The SFH4 galaxies have contributed a factor-of-three less mass, but most of this is the result of their lower frequency of 8\%, so this means that all 4 types contribute a similar amount of stellar mass from $6<z<12$. Finally, we plot the integrated stellar mass for this special epoch, which reaches $4\times10^{11}$\Msun in this volume at $z=6$, and is growing at a rate of $\sim$500\Msuns Myr$^{-1}$. 

We hope that both the rates and manner of star formation in these youngest of galaxies will provide the first meaningful constraints for numerical modeling studies of the evolution of the universe at the end of the first billion years of cosmic history.

\section{SFHs Across Space}
\label{sec:SFHS Space}
SFHs are known to have a strong spatial variance, in the sense that different kinds of galaxies (for example, ellipticals as opposed to spirals) are found to dominate in different environments \citep{Dressler1980}, and the different SFHs of these have clearly differed greatly.  Recognizing this, we looked at the spatial distributions to see if galaxy `environment' could be connected to the four types of SFHs.  Figure~\ref{fig:skymaps} shows the distribution on the sky of the four SFHs (again with the same color encoding).  There is obvious large-scale structure in the z1, z2, and z3 maps, especially in the lowest redshift $z\sim6$ map, where the contrast between large voids and  substantial clustering is strong. This is probably both a result of the growing clustering with epoch, but also because our much larger sample makes any contrast more discernible.  Still, there is a strong impression of substantial large-scale structure over the period $6<z<10$ --- covering most of the epoch of reionization.  A map of only the 23 galaxies in the z4 sample is not useful, of course, but we have made another view of the $z=11-12$ universe that we discuss below.

It is not surprising that there are no \emph{visual} spatial distribution differences in the four SFHs, but it is reasonable to expect that correlations of SFH types with local-density, or with nearest-neighbor distance, might provide some insight into whether the environments of these galaxies influence their SFHs.  In Figure 11 (\emph Appendix A) we show histograms of local-density and nearest-neighbor-distance for the $z\sim$, $z\sim7$, and $z\sim8-z9$ redshift ranges (aka, z1, z2, z3). The left three panels show density ranges that vary from a few to tens of galaxies arcmin$^{-2}$.  As with Figure~\ref{fig:SFHs_across_time}, the most notable feature of these diagrams are ups-and-downs likely associated with density fluctuations, with no clean separation by type. Perhaps there is a slight preference for the longer SFHs to be in denser regions or closer to their neighbors, but nothing clear enough to be helpful.

\begin{figure*}
\centerline{
\includegraphics[width=7.4in, angle=0]
{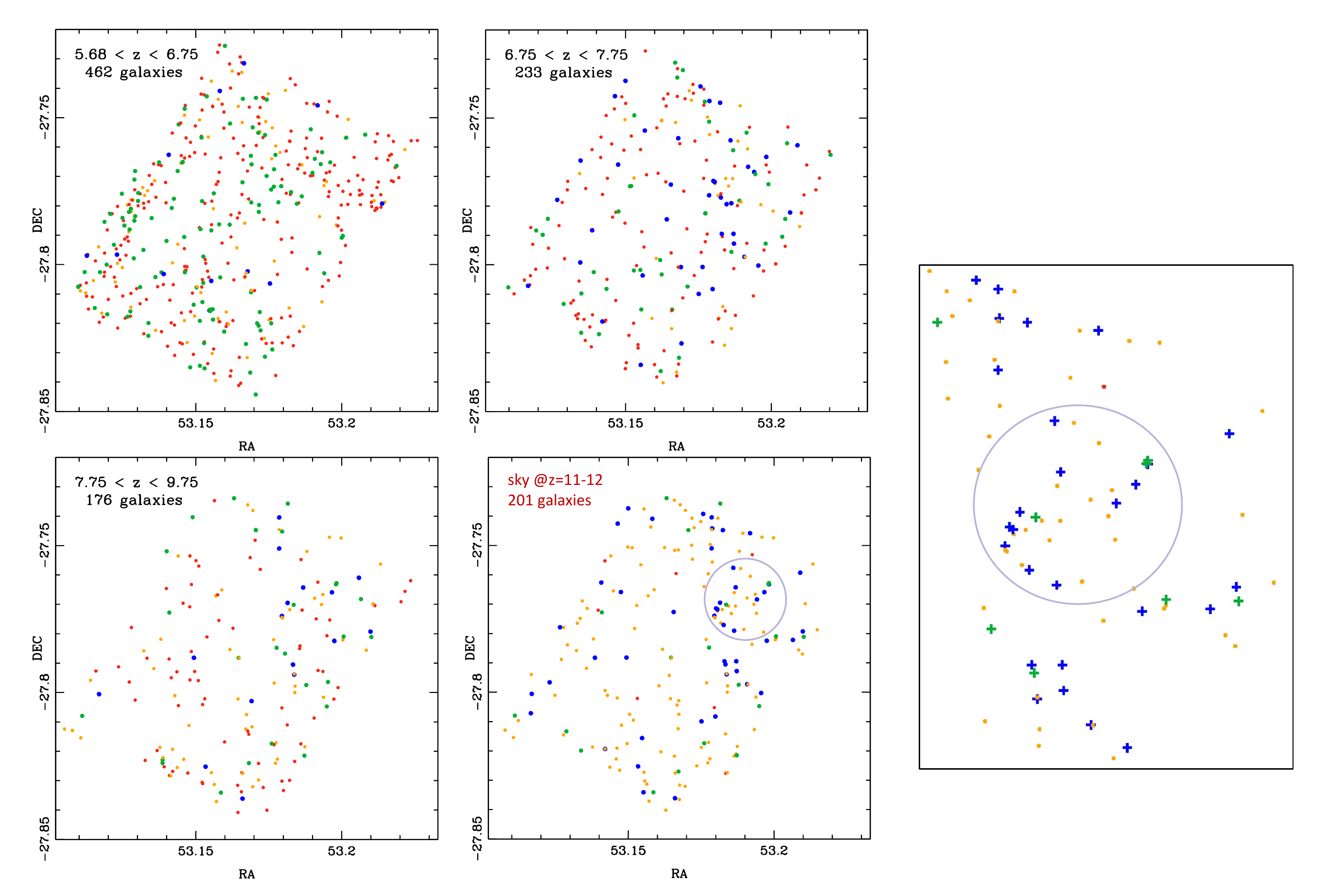}
}
\caption{The distribution on the sky for the three redshifts ranges z1, z2, \& z3, along with a map of galaxies found by \SEDz to have had `first star formation' at $z\sim11-12$ (lower right --- see text).  The four square figures essentially cover the full period of reionization.  Large-scale structure, in the form of large voids and swaths of higher galaxy density, is evident in each map, for the z1, z2, \& z3 samples.  The SFH of each galaxy is represented by color:  SFH1: burst = \emph{red}; SFH2: stochastic (multiburst) = \emph{orange};  SFH3: 3-contiguous epochs of star formation = \emph{green}; SFH4: long, continuous star formation = \emph{blue}.   However, there are no obvious correlations between location with respect to other galaxies of different SFH types visible from these maps, so if these exist, they must hold for higher density contrasts.  The bottom-right square map shows just such a higher concentration of galaxies --- those with `first star formation' at $z=11-12$ (this tight group is also visible in the panel above), with the surprisingly clear result that the rarest of our SFHs --- long and continuous --- are highly represented compared to the study sample.  An enlargement of the area ($\sim$2 arcmin) appears on the right.
} 
\label{fig:skymaps}

\end{figure*}

However, in the sky maps of Figure~\ref{fig:skymaps}, the fourth panel may hold an important clue about the role of environment.  Here we have done something new, based on the ability of \SEDz to identify prior star formation from galaxies observed at later epochs --- galaxy star-formation histories. We made this map by choosing galaxies for which \emph{substantial} star formation has been detected at $z=11$ or $z=12$,\footnote{We consider these epochs to be effectively indistinguishable} and in doing so, selected 201 galaxies of our sample of 894 that had their \emph{first epoch} of star formation at that time: we plot in the fourth panel what a sky with galaxies forming stars at z=11-12 might have looked like.

Remarkably, we see a relatively tight galaxy group of 30 members in an area less than $\sim$2 arcmin$^{-2}$. Eleven of those, 37\%, are SFH4 ---long, continuous SFHs, only 8\% of the full sample in this study.    We claim that this detection of many long SFHs `spatially,'  while not statistically conclusive, is a unique data point in the search for an environmental dependence of SFH type.  Furthermore, this distribution further confirms that \SEDz actually works: there is no other explanation for how these two quantities --- position on the sky and SFH --- could be well-correlated.  If  the star formation at $z=11-12$ \SEDz detected were bogus, these galaixes would spread randomly across the field.\footnote{\citep{Dressler2018} found a similar spatial dependence for \emph{Late Bloomers} and argued that their affinity for other late bloomers confirmed the legitimacy of their unconventional SFHs} The clustering, and also what appears to be large-scale structure for the full sample of galaxies, confirms these `earliest' SFHs.  In addition, above and below the circled area, the density remains high and appears to contain a higher fraction of SFH4 galaxies, compared to the much larger sample of the left side of the figure.

Subsequently, we noticed that this this higher-galaxy-density region also shows in the upper right panel, $6.75<z< 7.75$. In fact, this galaxy group had already been independently discovered by members of the \emph{JADES} team \citep{Endsley2023}, at an \emph{observed} redshift of $z\sim7.5$, further confirming their long SFHs.  Even more members are observed at the later epoch.

This fortunate `feature' of  the \emph{GOODS-S} field called to mind that, for the original morphology-density relationship, \citet{Dressler1980} and \citet{PostmanGeller} found no gradient in morphological type beyond the rich-cluster boundaries, that is, no slowly changing populations beyond the effective radius of the cluster. This and other evidence convinced Dressler (see Section IV-C) that the origin of the morphology-density relation lay not in the late-time evolution of the cluster or its galaxies.  Rather, it pointed to an environmental dependence arising at the epoch of galaxy birth: the different morphological types were foretold by their birth environment, a kind of \emph{early nurture} that is in fact \emph{nature}.

Soon there will be more deep-field imaging like \emph{GOODS-S} and more groups will be found at --- incredibly --- at $z\sim12$.  This result, if confirmed by many other cases, might suggest a rather obvious conclusion about the different SFHs we have found here: burst histories are most common in the equivalent of the lower density ``field'' of the modern universe, probably the result of stochastic merger events that reflect the sparser environment.  In this picture, long SHFs are destined for the richer, denser environments of the future. They would be galaxies that, unlike the ``bursts," were built up in a more orderly series of accretion events and minor mergers.

\section{Writing Chapter 3: The Age of Starbursts?}
\label{sec:The Age of Starbursts}

{Observations of the early universe with the incomparable \emph{JWST} already have already had a profound effect on our ideas about how the first galaxies were born and how they grew.  In scarcely more than a year, hundreds of studies have feasted on galaxy samples that are large, deep, various, and multiplexed.  They have been probed with diverse modes of cameras and spectrographs that are unprecedented in a space telescope, offering factors of hundreds to thousands greater sensitivity in the near-to-mid-IR: the early universe is viewed anew.}

{Our study is but one of multitudes focused on the rise of stars and galaxies, the elements that redefined the universe at the very epoch we now explore --- for the first time.  So far, this wealth of \JWST data tells a new story: that the birth of galaxies appears to be a volatile affair.   Specifically, a central theme has been that ``starbursts" are prevalent in this first billion years, or at the very least, that they play a critical role in understanding what see.  Before comparing the present work to a small sample of these papers, it is important to reiterate why --- for a majority of galaxies in our \emph{GOODS-S} field --- we identify their SFH as ``bursts."  Our study has particular significance because it was the first to trace the buildup of the stellar mass in the first galaxies \citep{Dressler2023}, and because it is the only one to rely solely on the light of main-sequence A-stars.  The mass-to-light ratios of these stars are thus accurately-known and independent of chemical composition.  Furthermore, a stellar population with an age of between 100 and 1000 Myr is not a dusty one: even while O \& B stars are still forming in HII regions, A-stars have already migrated into near-dustless environments.}

{Our study of 894 newborn galaxies at $6<z<12$ has shown that their star formation was predominately in bursts: we have called these histories ``starbursts"(single) and ``stochastic" (multiple). Because of the methodology of this study and the limitations of SEDs with 7 broad-bands fluxes, we actually know little about the nature of these bursts that we find dominating the SFHs of the first galaxies. For these, we observe primarily, almost exclusively, the light of A-stars whose ages range between $\sim$50--150 Myr. It is likely that the mass of stars \emph{born during} the full epoch was much less than that born in the first $\ls$50 Myr of explosive star formation that preceded it. In other words, most of the observed A-stars came from a shorter period that started earlier in the epoch. By this reasoning, we suggest that star formation rates of 10-30 \MsunYr\ {characterized the dust-enshrouded phase we have imagined, and it is during this phase that the $\sim$50--120 Myr-old A-stars that define our SEDs were born. We can estimate the SFRs of the CSF at OE and imagine them to be typical of the SFR over the whole epoch: that might only be 1-10} \MsunYr\ {(see, for example \citet{2019ApJ...881...71E, 2019ApJ...884..133F, 2023MNRAS.526.1512R}), but we will be hard-pressed to make a SED measurement, or any other we can think of, that can recover that \emph{history} of the SFR with a resolution of tens of megayears --- for any given object.}}\footnote{Alternatively, spectroscopic measurements of emission-line spectra for these kinds of galaxies could be used in a statistical way to estimate the range of SFRs and the level of ``burstiness" for the sample, and by comparing SFRs from UV flux to those from \Ha\ for large samples.}  

{It is more accurate, then, to say that our study has concluded that starbursts dominate not because we see signatures of} $10^8-10^9$\Msun of stars \emph{forming over an epoch}, {but because we see a population of A-stars in that epoch that are limited to ages less than $\ls$150 Myr. Most likely, they were born in a much shorter period within that epoch.  What defines our result that ``starbursts dominate" is that we see \emph{no} A-stars \emph{older} than, say, 150 Myr, that is, \emph{from a prior epoch.} That's why we call these SFHs bursts, not because we are seeing the burst (although we may be, in some cases), but because the only stars we do see are \emph{young.}
}

{We think that the principal benefit of our observations of the prominent role of starbursts will be to help inform theory, and numerical simulations models, about the growth of the baryonic component of the universe that winds up as \emph{galaxies}. The prevalence of bursts of star formation in the first billion years should be influential in guiding theoretical work to understand galaxy growth in a dynamic environment. It also seems that there are environmental factors that might send galaxies down one path rather than another that could help constrain numerical simulations.}

{Early results from \JWST have focused on a the large population of bright galaxies, particularly at $z>10$, \eg{\citet{10.1093/mnras/stad471}, \citet{Harikan+2024}, and \citet{2023ApJ...946L..13F}}, suggesting that the luminosity function of the first-galaxies evolved less slowly with redshift.\footnote{That is, the luminosity function not declining as rapidly as pre-\JWST (\emph{HST}) observations and derived predictions.}  Observations of \emph{brighter-than-expected}, \emph{more-numerous-than-expected}, or \emph{evolving-less-rapidly-than-expected}" galaxies have promoted ``burstiness" as a way to reconcile these findings (\eg{\citet{looser2023jades}}) with previous models (\eg{\citet{Wilkins2022}}).  However, those galaxies are higher redshift, and/or rarer (and more massive?) than those studied here, so our sample adds little to that discussion.}

{On the other hand, \citet{2023MNRAS.526.2665S} and \citet{2023ApJ...955L..35S} have in particular suggested that lower-mass galaxies, with an abundance and mass predicted pre-\JWST models, have been elevated in luminosity by bursts, specifically that their \emph{light-to-mass ratios} have risen through substantial bursts of star formation.  While again, those samples are not a comparable our own, we can by analogy question this explanation.  Our study finds a high level of `burstiness' through measurements of \emph{stellar mass} --- the mass of A-stars from the SEDs.  Likewise, the $z>10$ `bright' examples may be cases of galaxies forming more \emph{stellar mass} --- in bursts.}

{In a study that also seeks to measure a change in SFHs over $6<z<12$, \citet{2023arXiv230915720C} use a very different method than our own to extract such information from SEDs.  Their analysis correlates SED-shape with position on 
\emph{the starforming-main-sequence} to characterize whether a galaxy's SFH is `stochastic,' like our definition --- bursty over multiple epochs, or `secular,' --- longer and steadier, like our SFH3 and SFH4 types. They posit a smooth transition from mostly-stochastic to mostly-secular at around $z\sim9$. Although our sample is relatively small beyond $z>9$, Figure~\ref{fig:SFHs_across_time} shows no clear transition at $z>9$, with a similar fraction of SFH1+SFH2 and SFH3+SFH4 down to $z\sim7$. There seems to be a strong upturn in the both SFH1 andd SFH3  below $z\sim7$ (see Figure~\ref{fig:SFHs_rough_dist} with the ratio between `stochastic' (SFH1) and `secular' (SFH3) basically unchanged.  Again, our methodology is more straightforward --- \SEDz actually derives SFHs from ages of A-star populations, but this important issue requires further study.}

{About this important matter, what seems clear is that the most of the youngest galaxies do not grow steadily in a calm, peaceful environment. Rather, their journey to what we today regard as ``galaxy-sized" may be more chaotic, or even explosive. Perhaps these are gas-rich mergers dominating in lower density regions, manifesting in ``one-  or two-event" growth spurts in the first billion years, reaching $\sim$$10^9$ \Msun (what we now call galaxy-sized) when this first phase completes.  {Importantly, however, by $z\sim2$ --- ``cosmic noon" --- these galaxies should grow by an order-of-magnitude in mass to reach a halfway point for \emph{L$^*$}.  It will be important, and challenging, to relate that later, more-easily observed universe to what we have witnessed for the earliest galaxies. These starburst galaxies at $6<z<12$ neither seem poised for another burst, nor prepared to settle into steadier star-formation histories. The simple fact that most of our 894 galaxies do not show star formation over many epochs precludes the notion that these bursty objects are `picking up' again before $z\sim6$. Perhaps the ``contiguous" SFH3 galaxies, common in our $z=6$ sample, are destined to become those L* galaxies.  We note that almost all show the signatures of continuing star formation at $z=6$, so perhaps they will grow into the most common galaxies of today, just beginning in earnest their journey to maturity. It is undeniable that a transition from `stochastic' to `secular'} (to use Ciesla \etal's term) \emph{did} occur, so a focused study on $4<z<6$ galaxies, by whatever means their SFHs can be characterized, seems a priority, and imperative for ``Chapter 3."}

{Finally, pursuing our suggestion that the starbursts we have observed were the consequence of intense, extreme-density, dust-enshrouded bursts of star formation, we hope an effort can be made to match up this population of growing galaxies to a coeval population of heavily-dust-obscured proto-galaxies that will come from \emph{JWST/MIRI} and perhaps \emph{ALMA}.  That our sample seems effectively dust-free strongly suggests that there is a `just-before' phase of tens of millions of years in which explosive star formation ignited in gas-rich protogalaxies, leading directly to the objects we have been studying.  Forming stars rapidly in these sub-kiloparsec volumes is a formidable challenge to our considerable knowledge of how stars form.  In particular, the feedback from the supernovae of such starbursts should blow things apart!}  

{We have taken note, with great interest, of work suggesting that such feedback can be suppressed if, at that time, supernovae of mass M $\gs10$ \Msun collapsed directly into black holes.} {\citet{2022MNRAS.513.2111R} have suggested that this extraordinary explanation is near `mandatory' to explain the multiple stellar populations observed for many globular clusters, traditionally thought to be limited to a single generation by supernovae-driven feedback.  It appears that the same challenge applies towards our understanding of the earliest galaxies.  We find first ``bursts'' of star formation with} stellar masses $\gs$10$^8$\Msun\ --- {a scaled-up version of the largest globular clusters --- similarly in need of a physics miracle (see \citet{2023MNRAS.523.3201D}) to avoid destruction before fulfillment.}

{Travelers on \emph{JWST} --- the ultimate spacetimeship, we seek answers to questions first asked a century ago.  Wonderfully, we are fortunate to be grasping this `once in a species' opportunity to know our origins.}


\clearpage

\begin{acknowledgments}

AD gratefully acknowledges the support of the \emph{NIRCam} team for the opportunity to contribute to the \emph{NIRCam} science program on the earliest galaxies, and for prior funding from the \emph{JWST/NIRCam} contract,  NAS5-02015.  The DC2 ``deep field" simulation was a Herculean task that provided the essential guidance for the program described here, and for the \emph{NIRCam} GTO program in general.

This work is based on observations made with the NASA/ESA/CSA James Webb Space Telescope, using imaging data processed by the \emph{NIRCam} project team in collaboration with \emph{JADES}, a extragalactic project team consisting of members in the \emph{NIRCam} and \emph{NIRSpec} teams. D.S., K.M., M.R. and A.D. have been supported by \emph{JWST/NIRCam} contract to the University of Arizona, NAS5-02015.

N. Bonaventura acknowledges support from The Cosmic Dawn Center (DAWN), funded by the Danish National Research Foundation under grant no.140.

K. Boyett acknowledges partial support by the Australian Research Council Centre of Excellence for All Sky Astrophysics in 3 Dimensions (ASTRO 3D), through project number CE170100013.

A. Bunker acknowledges funding from the ``FirstGalaxies" Advanced Grant from the European Research Council (ERC) under the European Union’s Horizon 2020 research and innovation programme (Grant agreement No. 789056)

S. Cariani acknowledges support by European Union’s HE ERC Starting Grant No. 101040227 - WINGS.

D. Eisenstein is supported as a Simons Investigator and by \emph(JWST/NIRCam) contract to the University of Arizona, NAS5-02015

R. Hausen acknowledges funding for this research from the Johns Hopkins University, Institute for Data Intensive Engineering and Science (IDIES)

\end{acknowledgments}

\appendix 

\section{SFHs vs Environment at \MakeLowercase{z} $>$ 6 }

\begin{figure*}
\centerline{
\includegraphics[width=6.3in, angle=0]
{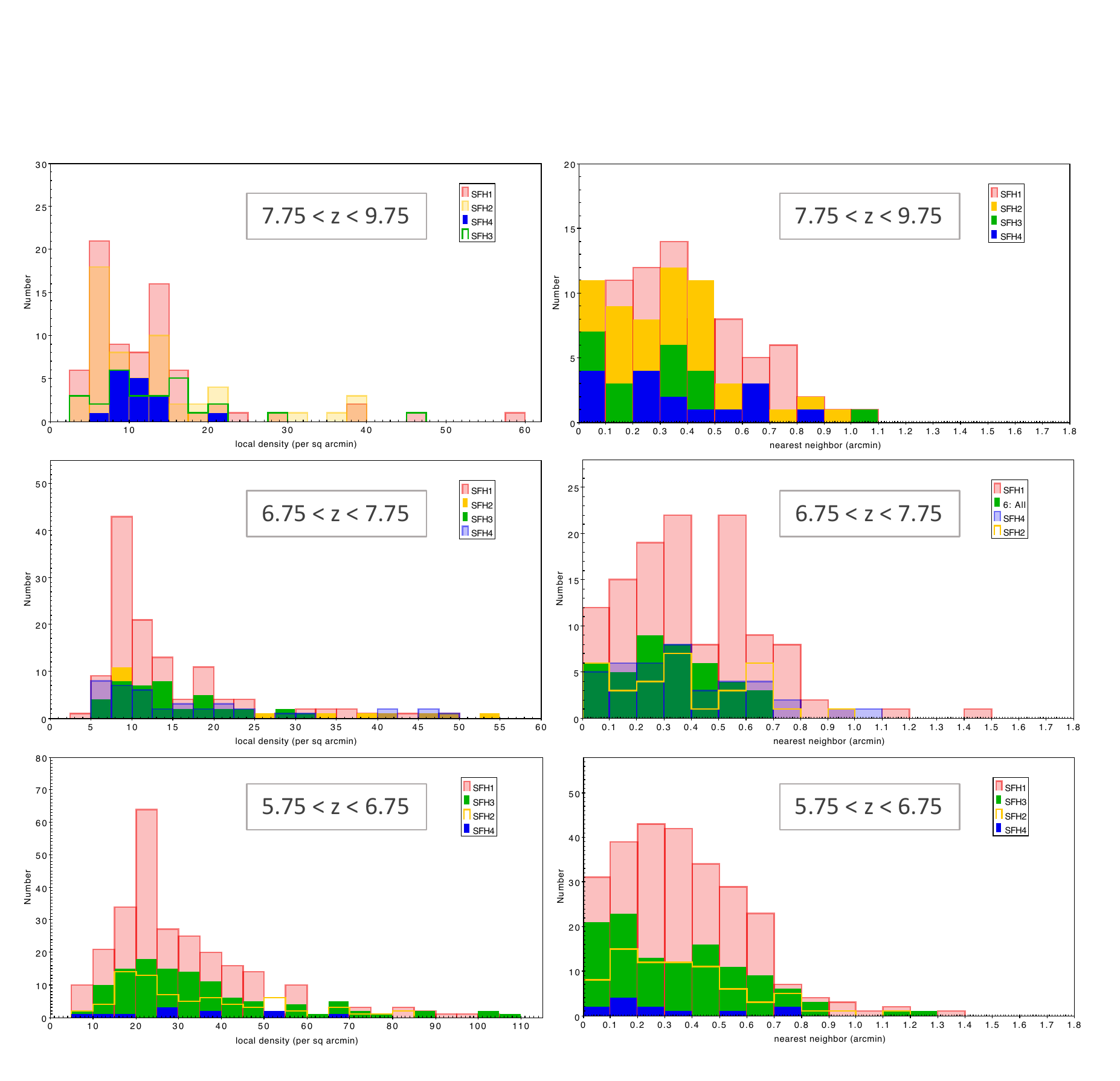}
}
\caption{Histograms of the incidence of the different SFH types with local density (\emph{left}) and as a function of nearest-neighbor-separation (\emph{right}).  No clear trends are apparent, that is, SFH types at this early time do not depend on local environment as they do after z$\sim$2   The three vertical panels are for the 3 redshift ranges, z $\approx$ 6, 7, and 8-10 (bottom to top).  The chief differences are in terms of scale, for example, the peak of both distributions shifts to lower density and to larger separation with increasing redshifts, as expected for a sample limited by apparent brightness.   Density contrasts in this \emph{GOODS-S} field are modest, but Figure~\ref{fig:skymaps} shows signs of a correlation of longer SFHs with higher density groups.
}
\label{density and nearest neighbor}
\end{figure*}

The histograms of Figure~\ref{density and nearest neighbor} show distributions of the four SFH types with local density --- calculated using the area of the 10 nearest neighbors galaxies, and distance to the single nearest neighbor galaxy.  The absence of any obvious trends between SFH type and environment seems at-odds with results for environmental dependencies later in cosmic history, for example, the obviously difference in SFHs of rapidly growing elliptical galaxies compared to the drawn-out SFHs of spirals like the Milky Way. However, from another perspective, the lack of environmental dependence is reminiscent of studies of the environment around rich clusters in the early 1980's \citep{Dressler1980, PostmanGeller} who found that a steep dependence of galaxy morphology with local density within the effective radius of a cluster did not continue to the lower density ``field" beyond.  In this study, Figure~\ref{fig:skymaps} shows a hint of the same behavior: in the region of higher density, a $z\sim$7 rich group, the long-SFH types are strongly represented, a correlation that is not expressed in lower-density surroundings.

\section{Stellar Population Templates of \emph{SEDz*} }

In this section we show samples of the stellar population templates used by \SEDz to characterize the SFHs of $5<z<12$ galaxies. Figure~\ref{fig:z12-burst-templates} plots the fluxes of 10 stellar population templates with a 10 Myr burst of star formation (at 1 \MsunYr = $10^7$\Msuns) at the start of epoch $z=12$, ``observed'' to evolve at epochs $z = 11...3$ --- later epochs without further star formation. For this study \SEDz uses 7 templates, for bursts starting at z = 11, 10...6.  The principal feature of this plot is that the templates are largely non-conformal, that is, not a conformal set of curves scaled by some parameter or set of parameters.  These ``vectors" describing stellar populations are different enough --- sufficiently \emph{orthonormal} --- that a least-squares combination of them is substantially `resolved' from any other combination.  This property allows \SEDz to `calculate' the history of a stellar population --- essentially, by vector algebra: finding the coefficients of the vector sum that best represent the observed SED.\footnote{Another way of thinking about these templates is to consider them as musical``chords," all playing the same 7 ``notes" (the wavelengths of the 7 wide-bands), but differing in the volume (flux) from templates to template.  Since all notes sound together, there are limited combinations available to fit the SED (the music), and there is no way to make changes in individual bands to improve the sound (fit). This is why the good SED fits found in this study --- with the restricted number of A-star-dominated templates from present-day populations --- provide a ``self-verification" of the method.}  What makes this particular application of the method potent is the non-conformal character of SEDs for stellar populations of ages $\tau<1~Gyr$ --- the templates covering the early universe for $z=12$ to $z=5$ whose light is dominated by main-sequence A-stars. The figure shows why, as has been known for half-a-century, finding the ages of stellar populations with stars only \emph{older} than 2 Gyr is, in practice, impossible: note how the templates $z=5, 4, 3$ are becoming a simple scaling of a single shape, as the universe reaches an age of 2 Gyr at $z\sim3$. 

\begin{figure*}
\centerline{
\includegraphics[width=4.8in, angle=0]{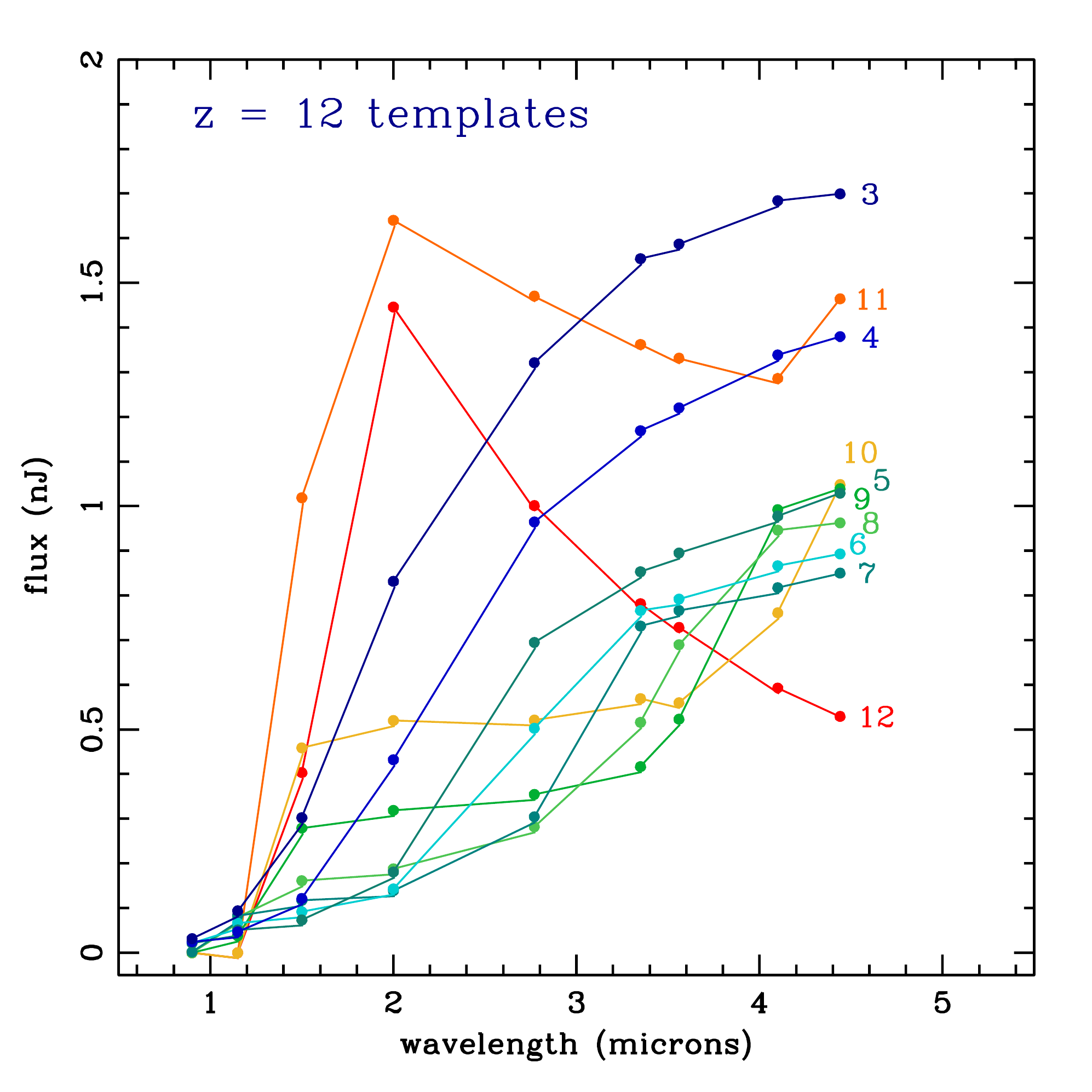}
}
\caption{Templates for a 10 Myr burst of star formation at $z=12$ evolved (aged and observed) at lower redshifts.  With good data (S/N $\gs$~7), the distinctive shapes of these templates allow \SEDz to make essentially unique SFH solutions.  This character changes for $z\le$~5 as the templates become scaled versions of each other, the reason that distinguishing stellar populations with $\tau>$2 Gyr are basically indistinguishable, a behavior attributable to the very similar \emph{giant branches} of older stars.
}
\label{fig:z12-burst-templates}
\end{figure*}

\begin{figure*}
\centerline{
\includegraphics[width=7.0in, angle=0]{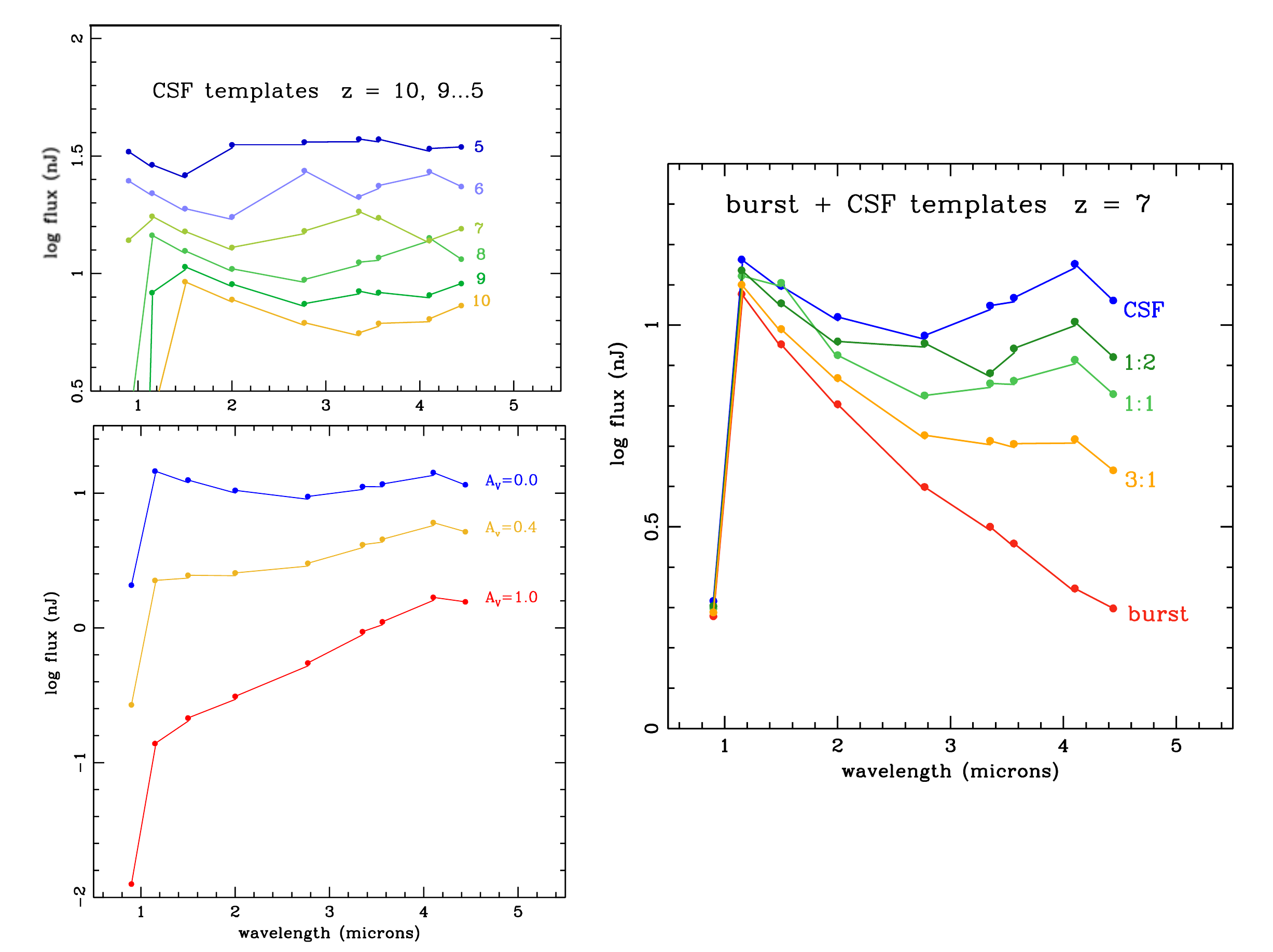}
}
\caption{\emph{top-left}: Templates for CSF (constant star formation) at epochs 5 through 10.  \emph{Right}: Combinations of burst and CSF templates at $z=7$.  Ratios are CSF-to-burst (see text).  \emph{left-bottom}: The affect of dust on the z=7 CSF template shown for $A_V$ = 0.4, a factor-of-two in the 1.15$\micron$ band, and $A_V$ = 1.0, a factor-of-ten, demonstrating the galaxies in the 894 sample analyzed here show little if any dust.  Even a 30\% decline across the SED from 4.44$\micron$ to 1.15$\micron$ --- equivalent to $A_V$ $\sim$ 0.1 --- would have adversely impacted the SEDz* fit compared to the no-dust solution.
}
\label{CSFandCSF+bursts_templates}
\end{figure*}

The signature of a burst of star formation is a very blue SED at that epoch, but for the subsequent epochs, the history of star formation \emph{within that epoch} is unresolved. Therefore, \SEDz accumulates the sum of bursts as the stellar masses of each: this is indistinguishable from continuous star formation over the prior epochs.

A final point of note is that \SEDz works because it is strongly constrained by the shape of each template, which means that variation of the coefficients in the maximum likelihood solution cannot either make or break the fit.  If these templates, made from present-epoch stars in our Galaxy, were not representative of stellar populations at $z>5$, this attempt to reproduce observed SEDs would fail badly. Quite the opposite is true.

The top-right panel of Figure~\ref{CSFandCSF+bursts_templates} shows templates for 6 templates, from $z=10$ to $z=5$, corresponding to the flux resulting from 1.0~\MsunYr\ of constant star formation (CSF) over that particular epoch.  Unlike bursts, there is no evolution of stellar population over subsequent epochs, because ongoing star formation can only be recorded in the SED from the epoch of observation. Prior epochs of CSF are indistinguishable from bursts of the same mass.  With the time resolution offered by broad-band SEDs, no additional information is available: there is no signature to distinguish CSF from a more complex behavior over the $\sim$100 Myr duration of each epoch.

Clearly, the distinguishing feature of the 6 CSF templates is that they are all flat --- as conformal as it gets --- in comparison to the burst templates.  The modulation that \emph{is} apparent comes from the Balmer break --- moving from $\sim2\micron$ at $z\sim6$ to $\sim3.5\micron$ at $z\sim10$,  and from the jaggedness of the SED from $3\micron$ to $5\micron$ --- due to \OIII and \Ha\ emission lines. This flatness, when combined with bursts of previous epochs, is responsible for much of `character' of the long SFHs.  The right panel of Figure~\ref{CSFandCSF+bursts_templates} shows that combining a burst, early in an epoch), with CSF at that same epoch, produces signatures that are found in many $z>5$ SEDs.  The ratios of 3:1 to 1:2 for the burst/CSF flux, shown in this case for $z\sim7$, are apparent in hundreds of the SEDs in our sample.   It is worth remembering that this combination is just equivalent to ongoing star formation that is declining, rather than constant, over the epoch.  

The bottom-left panel shows the  attenuation in the SED expected for different values of $A_V$, as described in Section~\ref{sec:special time}.

\section{Do Emission Lines Impact Derivations of SFH with \emph{SEDz*} ? }

In general, the influence of emission lines in photometric studies using broad bands, like this one, are only significant if the line-to-continuum ratio is very large.  For example, a moderate-resolution spectrograph (with resolutions of 100s) is effective at detecting even weak lines only when the continuum level is low, and that is true only for young stellar populations, $\tau\le20$ Myr.  In contrast, broad photometric bands like the ones used in this study select against such populations because the continuum flux is weak (the objects are faint). 

Because our detections of faint galaxies rely on the sensitivity to small fluxes, the equivalent width of an emission line must be enormous, that is, a low continuum flux.  Since the ``integer epochs” of this study cover $\sim$100 Myr of cosmic history, any object selected through broadband photometry will by necessity require a large burst of star formation compared to the stellar mass generated over the epoch. Thus, the contribution to the flux from the youngest populations, through emission lines, should be modest.

Figure 14 verifies that by selecting cases of relatively strong star emission lines (the right-side examples) and comparing to SEDs on the left with little evidence of star formation.  Our SEDs are made up of 7 broad bands (filled circles) and the two medium bands, F335M and F410M (open circles, not used in the \SEDz fitting). In particular, these and the F356W and F444W broad-bands are sampling the \OIII and \Ha\ lines over the redshift range $6<z<9.$ 

The medium bands often provide good evidence of emission because their width is $\sim$40\% of the broad bands, so they can also --- in cases that exhibit emission --- establish the continuum level at that color, sometimes considerably below that of the broad bands.\footnote{It is significant, though, that none of our SEDs show a continuum detection near zero, as would be the case for a pure emission-line spectrum, consistent with our claim that this is not possible with broad bands and long epochs.}  The close agreement in flux of both broad and narrow bands in the top and bottom left examples shows that there is little or no detected emission in these cases, which is common in our 894-galaxy sample. Note how the levels of F356W and F444W are close. The middle left SED shows moderate emission in F444W and detection of the continuum in F335M, and here F444W appears to be elevated above F356W by \OIII emission.  The case in the top right is a stronger example: strong detected emission in F356W and F335W, also \OIII, at this lower redshift.  

When compared to the flat blue (CSF) templates on the left, and purple on the right (excluding middle right), these SEDs all show a substantial rise in flux over the F356--F444 region.  What is the contribution of emission lines to this rise?  

For most of our sample, the answer is: little to none. This is because the CSF templates \emph{include} the emission line fluxes for the appropriate stellar population, for both the broad and narrow bands. As explained in \citet{Stark2013} the SEDs of these templates include hydrogen lines \citep{Robertson2010}, based on Osterbrock tabulations and case B recombination) and metallic lines (Z = 0.2 Z$\odot$, likely appropriate for our high-z galaxies) and continuum radiation, calculated and described by \citet{Stark2013}.  Of course, the strength of these lines should vary from object to object, and perhaps systematically from $z\sim0$ to $z>5$, but examination of dozens of our sample suggests that such variations, though present, are smaller than the effect of ``line vs. no-line."  However, the existence of variation, as well as not knowing the redshift well enough to place the emission lines accurately, prevent us from making detailed arguments about whether these low-z line ratios are in fact a good match for very young galaxies.

Knowing that the emission lines are appropriately included in the templates we use, is it easier to understand the examples we show here.  Since emission is included, the F356W and F444W fluxes of the upper and lower left SEDs show no emission, indeed, if their `elevation' from a flat SED were due to emission, the narrow bands would show higher, not at the same level.  The middle-left example does show elevation of F444W compared to its more typical closeness to the level of F356W, but the \SEDz model passes through both points because \emph{emission is included.}  (This indicates that the 'present-epoch' templates are appropriate.) The same goes for the top right example  with strong emission in F335M.  In this case and most others, the continuum level of F410M, compared to the elevated value of F356W, comes in at the proper ratio of 2.5:1.

\begin{figure*}
\centerline{
\includegraphics[width=6.7 in, angle=0]{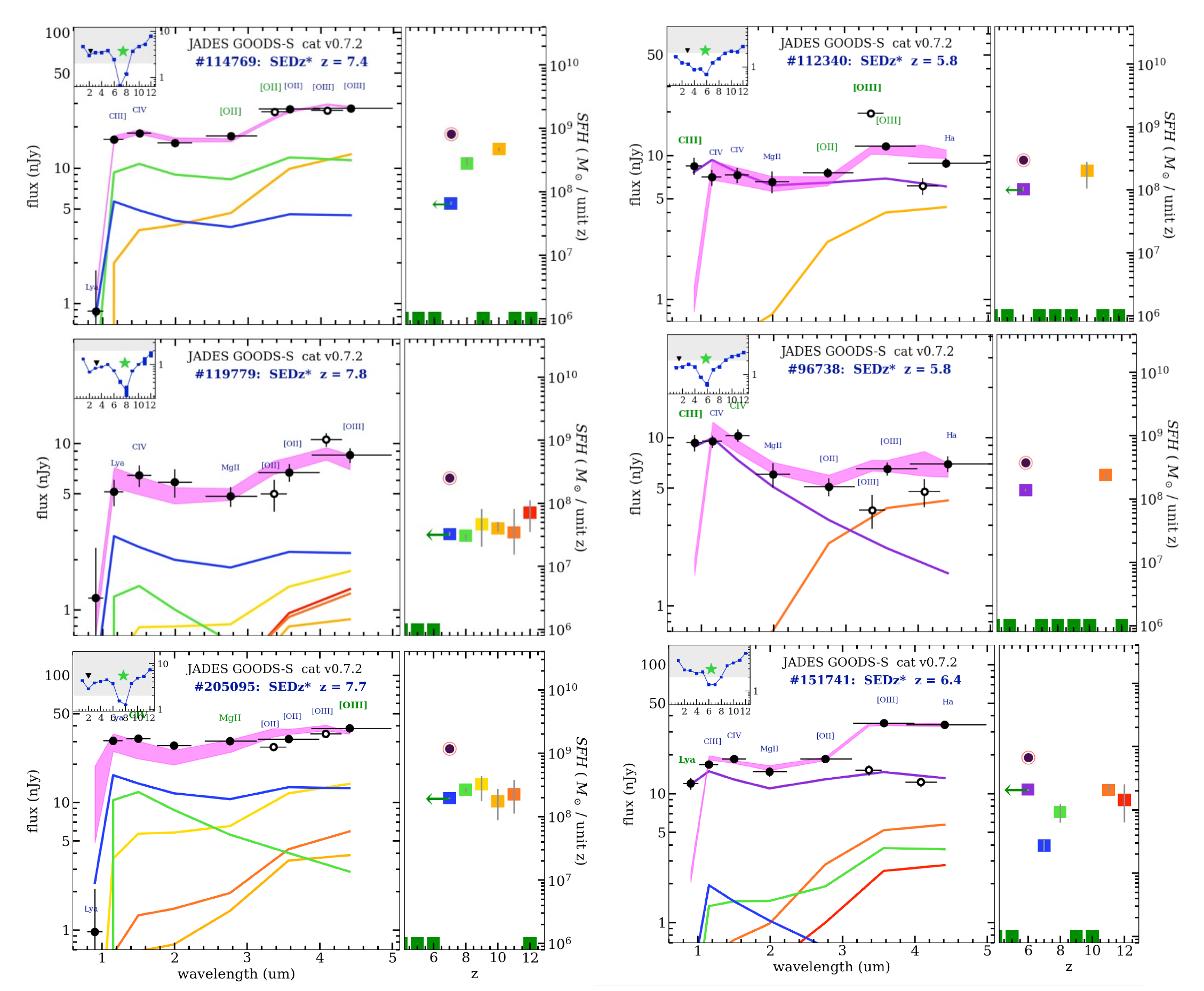}
}
\caption{These \SEDz plots address the influence of emission lines on SFHs derived in this study.  The SFHs on top and bottom left have little or no emission, as shown by the fact that the medium bands (2.5 times smaller bandwidth than broad bands) share the same flux level as the broad bands. The middle left shows a modest level of emission --- elevation of F410M over F444W, and the top right evidence for moderate emission.  The fact that the stellar population templates include emission-line flux representative of typical star-forming-regions is obvious from the middle and bottom right examples, demonstrating that the ``rise of the red continuum" in these 6 cases must be due to flux from older stellar populations, unless significant contributions from emission only \OIII and/or \Ha are \emph{substantially stronger than for present-day star-formation regions.} 
\label{fig:emission_no-emission}}
\end{figure*}

In the middle and bottom right examples we again see the \SEDz model passing through F356W and F444W, the emission --- here \OIII and \Ha\ --- is \emph{included}. The continuum below is sampled in each by both medium bands. Here we also see that the continuum level is provided by a single, older, $z=11$ population, which is boosting the level of F356W and F444 to give the SED its distinctive shape, one that is \emph{inconsistent with any single population}. For the bottom-right example the continuum is matched by $z=11-12$ and $z=8$ flux to complete the fit. The only way emission could contribute significantly would be if \OIII\ and \Ha\ emission were substantially stronger in these regions of star formation in $z>6$ galaxies than in  galaxies of today.

To look for galaxies with stronger-than-present-day emission, we examined SEDs to find cases where the points indicating emission had fluxes systematically higher than the values of the five bands F090W, F115W, F150W, F200W, and F277W.  There were few among the 690 cases of the z1 and z2 samples, $z<7.75$, but 50 of the 128 galaxies in the redshift range $7.75<z<9.75$ showed conspicuous \OIII emission in the F444 band.  Among these, 28 had SEDs that rose steadily through F356W, which \SEDz attributed to earlier star formation, $z\ge10$. For the remaining 22 the lack of such a clear trend suggested that these were, in  fact, cases of \OIII emission much stronger than represented by the templates. For such cases the ``earlier" stellar mass found by \SEDz was incorrect.  By examining the $z\ge10$ contribution, we determined that for such galaxies total stellar mass had been overestimated by factors of 1.5 to 3.5.  This is important for future \SEDz studies. At the same time, it does not alter the conclusions of this study, in fact, it moves some SFH2 types (an early and late burst) back to the SFH1 category, reaffirming our principal result of the prominence of starbursts in the early universe.

Since these strong emission cases are a small part of the study, this result, added to the cases with little-or-no emission, confirms  that ``red rise" we find for $\sim$100 galaxies  is flux from older stellar populations, in other words, \SEDz star formation histories confirmed.

\clearpage


\bibliography{main}

\begin{thebibliography}{}
\expandafter\ifx\csname natexlab\endcsname\relax\def\natexlab#1{#1}\fi
\providecommand{\url}[1]{\href{#1}{#1}}
\providecommand{\dodoi}[1]{doi:~\href{http://doi.org/#1}{\nolinkurl{#1}}}
\providecommand{\doeprint}[1]{\href{http://ascl.net/#1}{\nolinkurl{http://ascl.net/#1}}}
\providecommand{\doarXiv}[1]{\href{https://arxiv.org/abs/#1}{\nolinkurl{https://arxiv.org/abs/#1}}}

\bibitem[{{Bruzual} \& {Charlot}(2003)}]{Bruzual2003}
{Bruzual}, G., \& {Charlot}, S. 2003, \mnras, 344, 1000, \dodoi{10.1046/j.1365-8711.2003.06897.x}

\bibitem[{{Ciesla} {et~al.}(2023){Ciesla}, {Elbaz}, {Ilbert}, {Buat}, {Magnelli}, {Narayanan}, {Daddi}, {G{\'o}mez-Guijarro}, \& {Arango-Toro}}]{2023arXiv230915720C}
{Ciesla}, L., {Elbaz}, D., {Ilbert}, O., {et~al.} 2023, arXiv e-prints, arXiv:2309.15720, \dodoi{10.48550/arXiv.2309.15720}

\bibitem[{{Couch} \& {Sharples}(1987)}]{Couch1987}
{Couch}, W.~J., \& {Sharples}, R.~M. 1987, \mnras, 229, 423, \dodoi{10.1093/mnras/229.3.423}

\bibitem[{{Curtis-Lake} {et~al.}(2023){Curtis-Lake}, {Carniani}, {Cameron}, {Charlot}, {Jakobsen}, {Maiolino}, {Bunker}, {Witstok}, {Smit}, {Chevallard}, {Willott}, {Ferruit}, {Arribas}, {Bonaventura}, {Curti}, {D'Eugenio}, {Franx}, {Giardino}, {Looser}, {L{\"u}tzgendorf}, {Maseda}, {Rawle}, {Rix}, {Rodr{\'\i}guez del Pino}, {{\"U}bler}, {Sirianni}, {Dressler}, {Egami}, {Eisenstein}, {Endsley}, {Hainline}, {Hausen}, {Johnson}, {Rieke}, {Robertson}, {Shivaei}, {Stark}, {Tacchella}, {Williams}, {Willmer}, {Bhatawdekar}, {Bowler}, {Boyett}, {Chen}, {de Graaff}, {Helton}, {Hviding}, {Jones}, {Kumari}, {Lyu}, {Nelson}, {Perna}, {Sandles}, {Saxena}, {Suess}, {Sun}, {Topping}, {Wallace}, \& {Whitler}}]{CurtisLake_2023}
{Curtis-Lake}, E., {Carniani}, S., {Cameron}, A., {et~al.} 2023, Nature Astronomy, 7, 622, \dodoi{10.1038/s41550-023-01918-w}

\bibitem[{{Dekel} {et~al.}(2023){Dekel}, {Sarkar}, {Birnboim}, {Mandelker}, \& {Li}}]{2023MNRAS.523.3201D}
{Dekel}, A., {Sarkar}, K.~C., {Birnboim}, Y., {Mandelker}, N., \& {Li}, Z. 2023, \mnras, 523, 3201, \dodoi{10.1093/mnras/stad1557}

\bibitem[{Donnan {et~al.}(2023)Donnan, McLeod, McLure, Dunlop, Carnall, Cullen, \& Magee}]{10.1093/mnras/stad471}
Donnan, C.~T., McLeod, D.~J., McLure, R.~J., {et~al.} 2023, Monthly Notices of the Royal Astronomical Society, 520, 4554, \dodoi{10.1093/mnras/stad471}

\bibitem[{{Dressler}(1980)}]{Dressler1980}
{Dressler}, A. 1980, \apj, 236, 351, \dodoi{10.1086/157753}

\bibitem[{{Dressler} \& {Gunn}(1983)}]{Dressler1983}
{Dressler}, A., \& {Gunn}, J.~E. 1983, \apj, 270, 7, \dodoi{10.1086/161093}

\bibitem[{{Dressler} {et~al.}(2018){Dressler}, {Kelson}, \& {Abramson}}]{Dressler2018}
{Dressler}, A., {Kelson}, D.~D., \& {Abramson}, L.~E. 2018, \apj, 869, 152, \dodoi{10.3847/1538-4357/aaedbe}

\bibitem[{{Dressler} {et~al.}(1996){Dressler}, {Brown}, {Davidsen}, {Ellis}, {Freedman}, {Green}, {Hauser}, {Kirshner}, {Kulkarni}, {Lilly}, {Margon}, {Porco}, {Richstone}, {Stockman}, {Thronson}, {Tonry}, {Truran}, \& {Weiler}}]{Dressler1986}
{Dressler}, A., {Brown}, R.~A., {Davidsen}, A.~F., {et~al.} 1996, {Exploration an the Search for Origins: A Vision for Ultraviolet-Optical-Infrared Space Astronomy}, Technical Report, HST \& Beyond Committee

\bibitem[{{Dressler} {et~al.}(2023){Dressler}, {Vulcani}, {Treu}, {Rieke}, {Burns}, {Calabr{\`o}}, {Bonchi}, {Castellano}, {Fontana}, {Leethochawalit}, {Mason}, {Merlin}, {Morishita}, {Paris}, {Bradac}, {Mercurio}, {Nanayakkara}, {Poggianti}, {Santini}, {Wang}, {Misselt}, {Stark}, \& {Willmer}}]{Dressler2023}
{Dressler}, A., {Vulcani}, B., {Treu}, T., {et~al.} 2023, \apjl, 947, L27, \dodoi{10.3847/2041-8213/ac9ebb}

\bibitem[{Eisenstein {et~al.}(2023)Eisenstein, Willott, Alberts, Arribas, Bonaventura, Bunker, Cameron, Carniani, Charlot, Curtis-Lake, D'Eugenio, Endsley, Ferruit, Giardino, Hainline, Hausen, Jakobsen, Johnson, Maiolino, Rieke, Rieke, Rix, Robertson, Stark, Tacchella, Williams, Willmer, Baker, Baum, Bhatawdekar, Boyett, Chen, Chevallard, Circosta, Curti, Danhaive, DeCoursey, de~Graaff, Dressler, Egami, Helton, Hviding, Ji, Jones, Kumari, Lützgendorf, Laseter, Looser, Lyu, Maseda, Nelson, Parlanti, Perna, Puskás, Rawle, Pino, Sandles, Saxena, Scholtz, Sharpe, Shivaei, Silcock, Simmonds, Skarbinski, Smit, Stone, Suess, Sun, Tang, Topping, Übler, Villanueva, Wallace, Whitler, Witstok, \& Woodrum}]{eisenstein2023overview}
Eisenstein, D.~J., Willott, C., Alberts, S., {et~al.} 2023, Overview of the JWST Advanced Deep Extragalactic Survey (JADES).
\newblock \doarXiv{2306.02465}

\bibitem[{{Emami} {et~al.}(2019){Emami}, {Siana}, {Weisz}, {Johnson}, {Ma}, \& {El-Badry}}]{2019ApJ...881...71E}
{Emami}, N., {Siana}, B., {Weisz}, D.~R., {et~al.} 2019, \apj, 881, 71, \dodoi{10.3847/1538-4357/ab211a}

\bibitem[{{Endsley} {et~al.}(2023){Endsley}, {Stark}, {Whitler}, {Topping}, {Chen}, {Plat}, {Chisholm}, \& {Charlot}}]{Endsley2023}
{Endsley}, R., {Stark}, D.~P., {Whitler}, L., {et~al.} 2023, \mnras, 524, 2312, \dodoi{10.1093/mnras/stad1919}

\bibitem[{{Faisst} {et~al.}(2019){Faisst}, {Capak}, {Emami}, {Tacchella}, \& {Larson}}]{2019ApJ...884..133F}
{Faisst}, A.~L., {Capak}, P.~L., {Emami}, N., {Tacchella}, S., \& {Larson}, K.~L. 2019, \apj, 884, 133, \dodoi{10.3847/1538-4357/ab425b}

\bibitem[{{Finkelstein} {et~al.}(2023){Finkelstein}, {Bagley}, {Ferguson}, {Wilkins}, {Kartaltepe}, {Papovich}, {Yung}, {Arrabal Haro}, {Behroozi}, {Dickinson}, {Kocevski}, {Koekemoer}, {Larson}, {Le Bail}, {Morales}, {P{\'e}rez-Gonz{\'a}lez}, {Burgarella}, {Dav{\'e}}, {Hirschmann}, {Somerville}, {Wuyts}, {Bromm}, {Casey}, {Fontana}, {Fujimoto}, {Gardner}, {Giavalisco}, {Grazian}, {Grogin}, {Hathi}, {Hutchison}, {Jha}, {Jogee}, {Kewley}, {Kirkpatrick}, {Long}, {Lotz}, {Pentericci}, {Pierel}, {Pirzkal}, {Ravindranath}, {Ryan}, {Trump}, {Yang}, {Bhatawdekar}, {Bisigello}, {Buat}, {Calabr{\`o}}, {Castellano}, {Cleri}, {Cooper}, {Croton}, {Daddi}, {Dekel}, {Elbaz}, {Franco}, {Gawiser}, {Holwerda}, {Huertas-Company}, {Jaskot}, {Leung}, {Lucas}, {Mobasher}, {Pandya}, {Tacchella}, {Weiner}, \& {Zavala}}]{2023ApJ...946L..13F}
{Finkelstein}, S.~L., {Bagley}, M.~B., {Ferguson}, H.~C., {et~al.} 2023, \apjl, 946, L13, \dodoi{10.3847/2041-8213/acade4}

\bibitem[{Hainline {et~al.}(2020)Hainline, Hviding, Rieke, Shivaei, Endsley, Curtis-Lake, Smit, Williams, Alberts, Boyett, Bunker, Egami, Maseda, Tacchella, \& Willmer}]{Hainline_2020}
Hainline, K.~N., Hviding, R.~E., Rieke, M., {et~al.} 2020, The Astrophysical Journal, 892, 125, \dodoi{10.3847/1538-4357/ab7dc3}

\bibitem[{{Harikane} {et~al.}(2024){Harikane}, {Nakajima}, {Ouchi}, {Umeda}, {Isobe}, {Ono}, {Xu}, \& {Zhang}}]{Harikan+2024}
{Harikane}, Y., {Nakajima}, K., {Ouchi}, M., {et~al.} 2024, \apj, 960, 56, \dodoi{10.3847/1538-4357/ad0b7e}

\bibitem[{{Kelson} {et~al.}(2014){Kelson}, {Williams}, {Dressler}, {McCarthy}, {Shectman}, {Mulchaey}, {Villanueva}, {Crane}, \& {Quadri}}]{Kelson2014}
{Kelson}, D.~D., {Williams}, R.~J., {Dressler}, A., {et~al.} 2014, \apj, 783, 110, \dodoi{10.1088/0004-637X/783/2/110}

\bibitem[{{Lawson} \& {Hanson}(1995)}]{Lawson1995}
{Lawson}, C., \& {Hanson}, R. 1995, Solving Least Squares Problems. Classics in Applied Mathematics (Philadelphia: SIAM)

\bibitem[{Looser {et~al.}(2023)Looser, D'Eugenio, Maiolino, Tacchella, Curti, Arribas, Baker, Baum, Bonaventura, Boyett, Bunker, Carniani, Charlot, Chevallard, Curtis-Lake, Danhaive, Eisenstein, de~Graaff, Hainline, Ji, Johnson, Kumari, Nelson, Parlanti, Rix, Robertson, Pino, Sandles, Scholtz, Smit, Stark, Übler, Williams, Willott, \& Witstok}]{looser2023jades}
Looser, T.~J., D'Eugenio, F., Maiolino, R., {et~al.} 2023, JADES: Differing assembly histories of galaxies -- Observational evidence for bursty SFHs and (mini-)quenching in the first billion years of the Universe.
\newblock \doarXiv{2306.02470}

\bibitem[{{Merlin} {et~al.}(2022){Merlin}, {Bonchi}, {Paris}, {Belfiori}, {Fontana}, {Castellano}, {Nonino}, {Polenta}, {Santini}, {Yang}, {Glazebrook}, {Treu}, {Roberts-Borsani}, {Trenti}, {Birrer}, {Brammer}, {Grillo}, {Calabr{\`o}}, {Marchesini}, {Mason}, {Mercurio}, {Morishita}, {Strait}, {Boyett}, {Leethochawalit}, {Nanayakkara}, {Vulcani}, {Bradac}, \& {Wang}}]{Merlin2022}
{Merlin}, E., {Bonchi}, A., {Paris}, D., {et~al.} 2022, \apjl, 938, L14, \dodoi{10.3847/2041-8213/ac8f93}

\bibitem[{{Morgan} \& {Keenan}(1973)}]{Morgan1973}
{Morgan}, W.~W., \& {Keenan}, P.~C. 1973, \araa, 11, 29, \dodoi{10.1146/annurev.aa.11.090173.000333}

\bibitem[{{Nanayakkara} {et~al.}(2022){Nanayakkara}, {Glazebrook}, {Jacobs}, {Bonchi}, {Castellano}, {Fontana}, {Mason}, {Merlin}, {Morishita}, {Paris}, {Trenti}, {Treu}, {Calabro}, {Boyett}, {Bradac}, {Leethochawalit}, {Marchesini}, {Santini}, {Strait}, {Vanzella}, {Vulcani}, {Wang}, \& {Yang}}]{Nanayakkara2022}
{Nanayakkara}, T., {Glazebrook}, K., {Jacobs}, C., {et~al.} 2022, arXiv e-prints, arXiv:2207.13860.
\newblock \doarXiv{2207.13860}

\bibitem[{{Postman} \& {Geller}(1984)}]{PostmanGeller}
{Postman}, M., \& {Geller}, M. 1984, \apj, 281, 95, \dodoi{10.1086/162078}

\bibitem[{{Renzini} {et~al.}(2022){Renzini}, {Marino}, \& {Milone}}]{2022MNRAS.513.2111R}
{Renzini}, A., {Marino}, A.~F., \& {Milone}, A.~P. 2022, \mnras, 513, 2111, \dodoi{10.1093/mnras/stac973}

\bibitem[{{Rezaee} {et~al.}(2023){Rezaee}, {Reddy}, {Topping}, {Shivaei}, {Shapley}, {Fetherolf}, {Kriek}, {Coil}, {Mobasher}, {Siana}, {Du}, {Khostovan}, {Weldon}, {Emami}, \& {Chartab}}]{2023MNRAS.526.1512R}
{Rezaee}, S., {Reddy}, N.~A., {Topping}, M.~W., {et~al.} 2023, \mnras, 526, 1512, \dodoi{10.1093/mnras/stad2842}

\bibitem[{{Rieke} {et~al.}(2023{\natexlab{a}}){Rieke}, {Kelly}, {Misselt}, {Stansberry}, {Boyer}, {Beatty}, {Egami}, {Florian}, {Greene}, {Hainline}, {Leisenring}, {Roellig}, {Schlawin}, {Sun}, {Tinnin}, {Williams}, {Willmer}, {Wilson}, {Clark}, {Rohrbach}, {Brooks}, {Canipe}, {Correnti}, {DiFelice}, {Gennaro}, {Girard}, {Hartig}, {Hilbert}, {Koekemoer}, {Nikolov}, {Pirzkal}, {Rest}, {Robberto}, {Sunnquist}, {Telfer}, {Wu}, {Ferry}, {Lewis}, {Baum}, {Beichman}, {Doyon}, {Dressler}, {Eisenstein}, {Ferrarese}, {Hodapp}, {Horner}, {Jaffe}, {Johnstone}, {Krist}, {Martin}, {McCarthy}, {Meyer}, {Rieke}, {Trauger}, \& {Young}}]{2023PASP..135b8001R}
{Rieke}, M.~J., {Kelly}, D.~M., {Misselt}, K., {et~al.} 2023{\natexlab{a}}, \pasp, 135, 028001, \dodoi{10.1088/1538-3873/acac53}

\bibitem[{{Rieke} {et~al.}(2023{\natexlab{b}}){Rieke}, {Robertson}, {Tacchella}, {Hainline}, {Johnson}, {Hausen}, {Ji}, {Willmer}, {Eisenstein}, {Pusk{\'a}s}, {Alberts}, {Arribas}, {Baker}, {Baum}, {Bhatawdekar}, {Bonaventura}, {Boyett}, {Bunker}, {Cameron}, {Carniani}, {Charlot}, {Chevallard}, {Chen}, {Curti}, {Curtis-Lake}, {Danhaive}, {DeCoursey}, {Dressler}, {Egami}, {Endsley}, {Helton}, {Hviding}, {Kumari}, {Looser}, {Lyu}, {Maiolino}, {Maseda}, {Nelson}, {Rieke}, {Rix}, {Sandles}, {Saxena}, {Sharpe}, {Shivaei}, {Skarbinski}, {Smit}, {Stark}, {Stone}, {Suess}, {Sun}, {Topping}, {{\"U}bler}, {Villanueva}, {Wallace}, {Williams}, {Willott}, {Whitler}, {Witstok}, \& {Woodrum}}]{2023ApJS..269...16R}
{Rieke}, M.~J., {Robertson}, B., {Tacchella}, S., {et~al.} 2023{\natexlab{b}}, \apjs, 269, 16, \dodoi{10.3847/1538-4365/acf44d}

\bibitem[{{Robertson} {et~al.}(2010){Robertson}, {Ellis}, {Dunlop}, {McLure}, \& {Stark}}]{Robertson2010}
{Robertson}, B.~E., {Ellis}, R.~S., {Dunlop}, J.~S., {McLure}, R.~J., \& {Stark}, D.~P. 2010, \nat, 468, 49, \dodoi{10.1038/nature09527}

\bibitem[{{Robertson} {et~al.}(2023){Robertson}, {Tacchella}, {Johnson}, {Hainline}, {Whitler}, {Eisenstein}, {Endsley}, {Rieke}, {Stark}, {Alberts}, {Dressler}, {Egami}, {Hausen}, {Rieke}, {Shivaei}, {Williams}, {Willmer}, {Arribas}, {Bonaventura}, {Bunker}, {Cameron}, {Carniani}, {Charlot}, {Chevallard}, {Curti}, {Curtis-Lake}, {D'Eugenio}, {Jakobsen}, {Looser}, {L{\"u}tzgendorf}, {Maiolino}, {Maseda}, {Rawle}, {Rix}, {Smit}, {{\"U}bler}, {Willott}, {Witstok}, {Baum}, {Bhatawdekar}, {Boyett}, {Chen}, {de Graaff}, {Florian}, {Helton}, {Hviding}, {Ji}, {Kumari}, {Lyu}, {Nelson}, {Sandles}, {Saxena}, {Suess}, {Sun}, {Topping}, \& {Wallace}}]{2023NatAs...7..611R}
{Robertson}, B.~E., {Tacchella}, S., {Johnson}, B.~D., {et~al.} 2023, Nature Astronomy, 7, 611, \dodoi{10.1038/s41550-023-01921-1}

\bibitem[{{Stark} {et~al.}(2013){Stark}, {Schenker}, {Ellis}, {Robertson}, {McLure}, \& {Dunlop}}]{Stark2013}
{Stark}, D.~P., {Schenker}, M.~A., {Ellis}, R., {et~al.} 2013, \apj, 763, 129, \dodoi{10.1088/0004-637X/763/2/129}

\bibitem[{{Sun} {et~al.}(2023{\natexlab{a}}){Sun}, {Faucher-Gigu{\`e}re}, {Hayward}, \& {Shen}}]{2023MNRAS.526.2665S}
{Sun}, G., {Faucher-Gigu{\`e}re}, C.-A., {Hayward}, C.~C., \& {Shen}, X. 2023{\natexlab{a}}, \mnras, 526, 2665, \dodoi{10.1093/mnras/stad2902}

\bibitem[{{Sun} {et~al.}(2023{\natexlab{b}}){Sun}, {Faucher-Gigu{\`e}re}, {Hayward}, {Shen}, {Wetzel}, \& {Cochrane}}]{2023ApJ...955L..35S}
{Sun}, G., {Faucher-Gigu{\`e}re}, C.-A., {Hayward}, C.~C., {et~al.} 2023{\natexlab{b}}, \apjl, 955, L35, \dodoi{10.3847/2041-8213/acf85a}

\bibitem[{{Tacchella} {et~al.}(2023){Tacchella}, {Eisenstein}, {Hainline}, {Johnson}, {Baker}, {Helton}, {Robertson}, {Suess}, {Chen}, {Nelson}, {Pusk{\'a}s}, {Sun}, {Alberts}, {Egami}, {Hausen}, {Rieke}, {Rieke}, {Shivaei}, {Williams}, {Willmer}, {Bunker}, {Cameron}, {Carniani}, {Charlot}, {Curti}, {Curtis-Lake}, {Looser}, {Maiolino}, {Maseda}, {Rawle}, {Rix}, {Smit}, {{\"U}bler}, {Willott}, {Witstok}, {Baum}, {Bhatawdekar}, {Boyett}, {Danhaive}, {de Graaff}, {Endsley}, {Ji}, {Lyu}, {Sandles}, {Saxena}, {Scholtz}, {Topping}, \& {Whitler}}]{Tacchella+2023}
{Tacchella}, S., {Eisenstein}, D.~J., {Hainline}, K., {et~al.} 2023, \apj, 952, 74, \dodoi{10.3847/1538-4357/acdbc6}

\bibitem[{{Treu} {et~al.}(2022{\natexlab{a}}){Treu}, {Roberts-Borsani}, {Bradac}, {Brammer}, {Fontana}, {Henry}, {Mason}, {Morishita}, {Pentericci}, {Wang}, {Acebron}, {Bagley}, {Bergamini}, {Belfiori}, {Bonchi}, {Boyett}, {Boutsia}, {Calabro}, {Caminha}, {Castellano}, {Dressler}, {Glazebrook}, {Grillo}, {Jacobs}, {Jones}, {Kelly}, {Leethochawalit}, {Malkan}, {Marchesini}, {Mascia}, {Mercurio}, {Merlin}, {Nanayakkara}, {Nonino}, {Paris}, {Poggianti}, {Rosati}, {Santini}, {Scarlata}, {Shipley}, {Strait}, {Trenti}, {Tubthong}, {Vanzella}, {Vulcani}, \& {Yang}}]{Treu2022a}
{Treu}, T., {Roberts-Borsani}, G., {Bradac}, M., {et~al.} 2022{\natexlab{a}}, ApJ, in press, arXiv:2206.07978.
\newblock \doarXiv{2206.07978}

\bibitem[{{Treu} {et~al.}(2022{\natexlab{b}}){Treu}, {Roberts-Borsani}, {Bradac}, {Brammer}, {Fontana}, {Henry}, {Mason}, {Morishita}, {Pentericci}, {Wang}, {Acebron}, {Bagley}, {Bergamini}, {Belfiori}, {Bonchi}, {Boyett}, {Boutsia}, {Calabr{\'o}}, {Caminha}, {Castellano}, {Dressler}, {Glazebrook}, {Grillo}, {Jacobs}, {Jones}, {Kelly}, {Leethochawalit}, {Malkan}, {Marchesini}, {Mascia}, {Mercurio}, {Merlin}, {Nanayakkara}, {Nonino}, {Paris}, {Poggianti}, {Rosati}, {Santini}, {Scarlata}, {Shipley}, {Strait}, {Trenti}, {Tubthong}, {Vanzella}, {Vulcani}, \& {Yang}}]{2022ApJ...935..110T}
---. 2022{\natexlab{b}}, \apj, 935, 110, \dodoi{10.3847/1538-4357/ac8158}

\bibitem[{Wilkins {et~al.}(2022)Wilkins, Vijayan, Lovell, Roper, Irodotou, Caruana, Seeyave, Kuusisto, Thomas, \& Parris}]{Wilkins2022}
Wilkins, S.~M., Vijayan, A.~P., Lovell, C.~C., {et~al.} 2022, Monthly Notices of the Royal Astronomical Society, 519, 3118–3128, \dodoi{10.1093/mnras/stac3280}

\bibitem[{{Williams} {et~al.}(2018){Williams}, {Curtis-Lake}, {Hainline}, {Chevallard}, {Robertson}, {Charlot}, {Endsley}, {Stark}, {Willmer}, {Alberts}, {Amorin}, {Arribas}, {Baum}, {Bunker}, {Carniani}, {Crandall}, {Egami}, {Eisenstein}, {Ferruit}, {Husemann}, {Maseda}, {Maiolino}, {Rawle}, {Rieke}, {Smit}, {Tacchella}, \& {Willott}}]{Williams2018}
{Williams}, C.~C., {Curtis-Lake}, E., {Hainline}, K.~N., {et~al.} 2018, \apjs, 236, 33, \dodoi{10.3847/1538-4365/aabcbb}

\end{thebibliography}
\bibliographystyle{aasjournal}

\end{document}